\documentclass[aps,prx,superscriptaddress,notitlepage,nopacs,amsmath,amstex,amssymb,citeautoscript,longbibliography,floatfix,letter]{revtex4-2}
\usepackage{graphicx} 
\usepackage{amsmath}
\usepackage{braket}
\usepackage{comment}
\usepackage[dvipsnames]{xcolor}
\usepackage[bookmarks=true,colorlinks,linkcolor=OrangeRed,urlcolor=NavyBlue,citecolor=RoyalBlue]{hyperref}
\usepackage{float}
\usepackage[normalem]{ulem}

\newcommand{\dab}{{\color{red}\downarrow}}
\newcommand{\uar}{{\uparrow}}

\begin{document}

\title{
Resonant false vacuum decay in two dimensions on a 4000-qubit quantum annealer
}

\author{Gregor Humar}
   \affiliation{Department of Complex Matter, Jožef Stefan Institute, Jamova 39, 1000 Ljubljana,       Slovenia}
   \affiliation{Department of Physics, Faculty of Mathematics and Physics, University of Ljubljana, Jadranska 19, 1000 Ljubljana, Slovenia}
\author{Jean-Yves Desaules}
   \affiliation{Institute of Science and Technology Austria (ISTA), Am Campus 1, 3400                   Klosterneuburg, Austria}
\author{Luka Pavešić}
    \affiliation{Department of Physics and Astronomy “G. Galilei”, University of Padova, via Marzolo 8, I-35131 Padova, Italy}
    \affiliation{National Institute for Nuclear Physics (INFN), Section of Padova, I-35131 Padova, Italy}
\author{Marko Ljubotina}
    \affiliation{Technical University of Munich, TUM School of Natural Sciences, Physics Department, James-Franck-Str. 1, 85748 Garching, Germany}
    \affiliation{Munich Center for Quantum Science and Technology (MCQST), Schellingstr. 4, München 80799, Germany}
\author{Zlatko Papi\'c}
   \affiliation{School of Physics and Astronomy, University of Leeds, Leeds LS2 9JT, UK}
\author{Kristel Michielsen}
	\affiliation{Jülich Supercomputing Centre, Forschungszentrum Jülich, D-52425 Jülich, Germany}
	\affiliation{Department of Computer Science, University of Cologne, 50931 Cologne, Germany }
\author{Jaka Vodeb}\email{jaka.vodeb@ijs.si}
   \affiliation{Department of Complex Matter, Jožef Stefan Institute, Jamova 39, 1000 Ljubljana,       Slovenia}
   \affiliation{CENN Nanocenter, Jamova 39, 1000 Ljubljana, Slovenia}
   \affiliation{Jülich Supercomputing Centre, Institute for Advanced Simulation, Forschungszentrum     Jülich, 52425 Jülich, Germany}

\date{\today}

\begin{abstract}
From cosmology to quantum matter, metastable states often decay through the nucleation and growth of competing domains, with false vacuum decay providing the paradigmatic example of this process. Here we demonstrate a distinct regime in which domain growth outpaces nucleation by orders of magnitude and is controlled by local resonance conditions. Using a programmable quantum annealer with more than 4000 qubits, we realize a two-dimensional quantum Ising model whose metastable spin-polarized state encodes a false vacuum. At a specific value of the longitudinal field, single-spin flips at the boundary of a seeded bubble become resonant, enabling kinetically constrained expansion. Combining experiment with tensor-network simulations and stochastic circuit modeling, we observe nearly ballistic growth of true-vacuum domains with sub-ballistic interface broadening, consistent with Kardar--Parisi--Zhang universality. Our results establish a growth-dominated regime of false vacuum decay and show how large-scale quantum simulation can access nonequilibrium metastable dynamics relevant to quantum field theory, cosmology, and strongly correlated matter.
\end{abstract}

\maketitle

\section*{Introduction}

Metastability is ubiquitous in many-body physics: a state can remain long-lived even when a lower-energy configuration is available, until fluctuations eventually trigger its decay. In quantum field theory and cosmology, the relaxation of a metastable vacuum through the nucleation of localized bubbles of the true vacuum is known as false-vacuum decay (FVD)~\cite{coleman1977fate,voloshin1974bubbles,coleman1980gravitational,linde1983decay}. More broadly, the same phenomenon underlies first-order quantum phase transitions~\cite{sachdev1999quantum} and the decay of metastable states in settings ranging from mesoscopic systems and quantum materials to reaction-rate theory, open quantum systems, and many-body quantum dynamics~\cite{iachello2004quantum,de2021colloquium,hanggi1990reaction,leggett1987dynamics,macieszczak2016towards,yin2025theory}.

The conventional description of FVD is critical-bubble theory: quantum or thermal fluctuations nucleate small domains of the stable phase, and the competition between bulk energy gain and interfacial tension determines whether such bubbles shrink or expand~\cite{coleman1977fate,voloshin1974bubbles,linde1983decay}. Once a bubble exceeds a critical size, its subsequent evolution is usually regarded as comparatively simple, with growth driven by the free-energy difference between the two phases. In this way, standard treatments separate the problem into \emph{nucleation} and \emph{growth}, with most of the conceptual emphasis placed on the former~\cite{guth1981inflationary,isidori2001metastability,degrassi2012higgs,caprini2020detecting,kuznetsov1997false,pavevsic2025scattering,gregory2014black,gelmini1994kinetics,enqvist1995induced,Guth1981cosmological,sarangi2009rapid,gleiser2005resonant,wang2026entanglement}. This leaves open the possibility of a qualitatively different regime in which post-nucleation growth itself becomes the nontrivial part of the dynamics, controlled by local resonant processes at the bubble boundary rather than by the smooth pressure-driven motion of an interface.

Motivated by the field-theoretic picture, lattice realizations of FVD in quantum spin systems were first explored theoretically~\cite{rutkevich1999decay,lagnese2021false,pomponio2022bloch,darbha2024false} and have recently become experimentally accessible. Analogs of FVD have now been realized in superfluids~\cite{zenesini2024false}, neutral-atom systems~\cite{cominotti2025observation,chao2025probing}, cold atoms~\cite{song2022realizing,zhu2024probing}, trapped ions~\cite{luo2025quantum}, quantum annealers~\cite{vodeb2025stirring}, and graphene multilayers~\cite{shavit2025ephemeral}. In one dimension, these studies probed decay-rate scaling, confinement effects, and the importance of properly preparing metastable states~\cite{lagnese2021false,pomponio2022bloch,chao2025probing}, revealing intrinsic lattice effects, such as quantized and resonant nucleation channels~\cite{vodeb2025stirring,chao2025probing}. By contrast, FVD in two dimensions remains far less understood, with recent numerical studies starting to access such regimes~\cite{borla2026microscopic} and Rydberg atom experiments observing cluster-nucleation phenomena~\cite{osterholz2025collective}. In particular, the real-time growth of resonant false-vacuum bubbles in two dimensions has largely remained unexplored.

Two dimensions are qualitatively different from one: bubble growth is no longer constrained to a line, but depends on the geometry of the interface and on the competition between area and perimeter. This makes two dimensions a natural setting 
to explore whether post-nucleation growth can itself become the essential part of FVD. Here, we address this question using a large-scale programmable quantum annealer that realizes the two-dimensional mixed-field Ising model in a metastable regime. We identify a special resonance at which single-spin flips at the boundary of an existing true-vacuum seed become energy conserving, causing bubble growth to outpace spontaneous nucleation by several orders of magnitude. Combining quantum simulation with tensor-network calculations and large-scale stochastic modeling, we show that this produces a distinct growth-dominated regime of FVD, characterized by ballistic expansion together with nontrivial interface broadening following the Kardar-Parisi-Zhang (KPZ) universality class. Our results uncover a new route to FVD in which local resonant processes govern the expansion of true-vacuum domains, and establish programmable quantum hardware as a controlled platform for studying FVD beyond the standard semiclassical paradigm.

\begin{figure}
    \centering
    \includegraphics[width=\columnwidth]{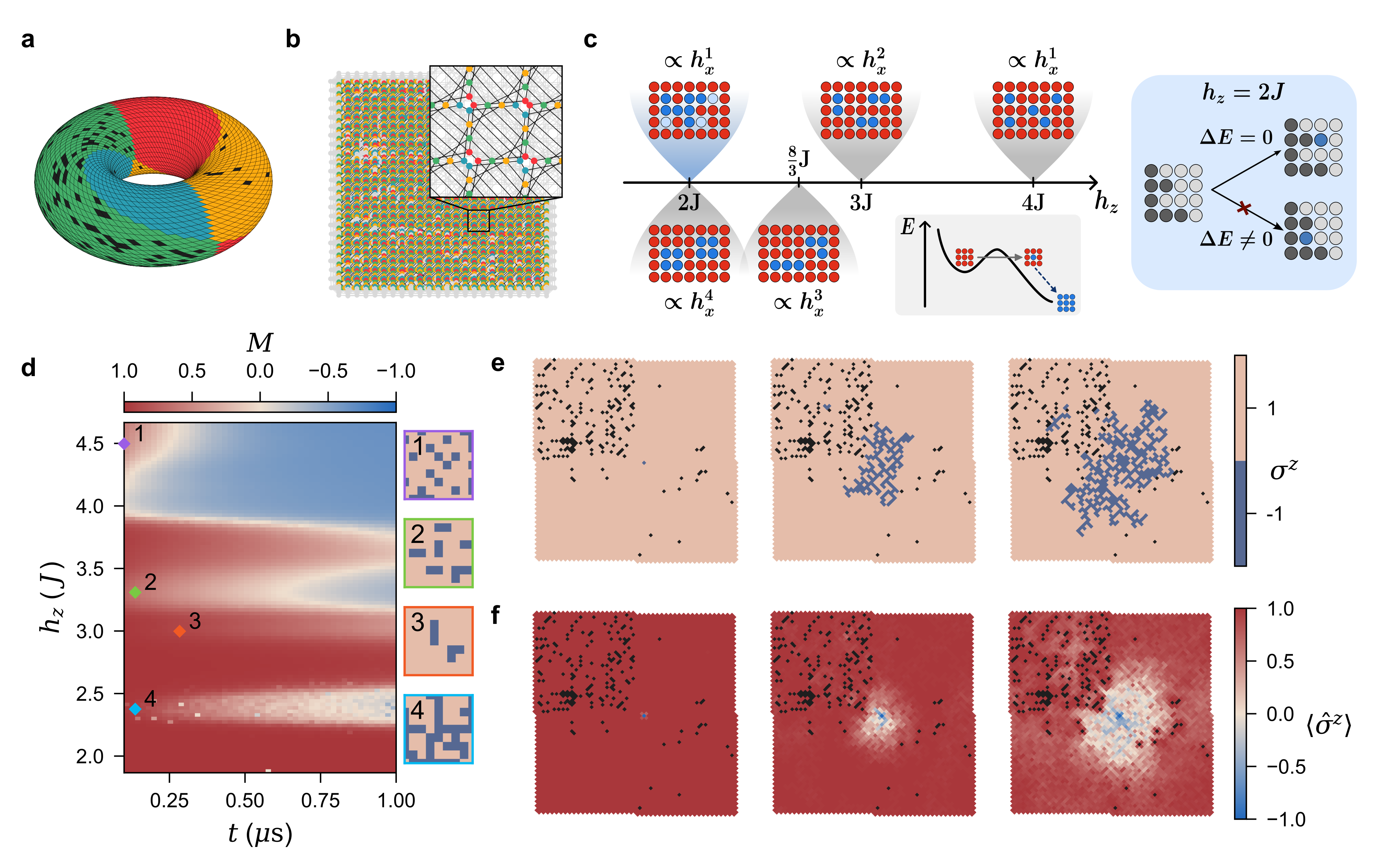}
    
\caption{\textbf{Two-dimensional resonant false-vacuum decay on a quantum annealer.}
\textbf{a,} Logical square lattice with periodic boundary conditions in both directions, realizing a torus topology. The colors indicate subsets of logical sites according to the physical-qubit groups used in the embedding shown in \textbf{b}. Black squares denote missing physical qubits on the processor, which are bypassed in the logical construction.
\textbf{b,} Embedding of the logical lattice on the quantum processing unit. The colored groups match those in \textbf{a} and are connected so as to implement the periodic two-dimensional geometry despite hardware defects.
\textbf{c,} Lowest resonances of the mixed-field Ising model as the longitudinal field is varied. The resonances correspond to the condition in Eq.~\ref{eq:res} where the domain-wall cost matches the energy gain from flipped spins. Shown are the one-, two-, three-, and four-spin resonances. The \(h_z=2J\) resonance is special because, in addition to four-spin nucleation, it also enables first-order resonant growth of an already existing bubble.
\textbf{d,} Experimental sweep of the longitudinal field on the annealer, showing the magnetization \(M\) as a function of \(h_z\) and evolution time \(t\) for the constant-\(h_z\) reverse-annealing protocol. The measured resonance structure agrees with the sequence shown in \textbf{c}; representative spin configurations at the numbered points are displayed on the right. The strongest resonances are the lowest-order ones, while the higher-order two- and three-spin resonances appear progressively weaker, consistent with the hierarchy of tunneling amplitudes. The resonant values of $h_z$ are slightly shifted with respect to expected values ($4J$, $3J$, $8J/3$, $2J$) due to slight miscalibration between the user specified values of $h_z$ and $J$ and actual device values, however, we characterized the resonances using the dominant bubble type we observed.
\textbf{e,} Three single-shot experimental configurations obtained from a deliberately prepared initial state containing a single flipped spin at the center of the system. Here, a time-dependent linear ramp of \(h_z(t)\) is used to pass through the \(h_z=2J\) resonance while \(h_x\) remains finite, thereby isolating the fast growth dynamics associated with the resonant interface process. A ramified fractal pattern spreads outward from the seed in a manner consistent with the local resonance rule. At later times, weak thermalization effects become visible through violations of the strict resonance constraint.
\textbf{f,} Time-averaged spin configurations obtained from 100 snapshots of the type shown in \textbf{e}, revealing the average spatial structure of the resonantly spreading bubble.}
    
    \label{fig:1}
\end{figure}

\section*{Quantum simulation of resonant false vacuum decay}

We study the mixed-field quantum Ising model
\begin{equation}
\hat{H}=-J\sum_{\langle i,j\rangle}\hat{\sigma}_i^z\hat{\sigma}_j^z
-h_x\sum_i \hat{\sigma}_i^x
-h_z\sum_i \hat{\sigma}_i^z,
\label{eq:fullmodel_rewritten}
\end{equation}
on a square lattice with periodic boundary conditions, in the regime \(J\gg h_x\). For \(h_z=0\), the model has two degenerate ferromagnetic ground states, \(\ket{\downarrow\downarrow\ldots\downarrow}\) and \(\ket{\uparrow\uparrow\ldots\uparrow}\). A finite longitudinal field lifts this degeneracy, so that one state becomes the true vacuum and the other a metastable false vacuum; here we take the all-down state as the true vacuum and the all-up state as the false vacuum. Figures~\ref{fig:1}a,b show the implementation of this model on the D-Wave Advantage2 quantum annealer. The logical lattice has the topology of a torus and is embedded into the Advantage2\_system1.6 and Advantage2\_system1.13 processors, using 4058 and 4044 out of 4593 and 4579 physical qubits, respectively.  The reverse-annealing protocol provides direct control over the time-dependent transverse field \(h_x(t)\), while the longitudinal field \(h_z(t)\) can be programmed independently, as described below. The exact details of the protocols are given in the Supplementary Information (SI)~\cite{SM}.

Domains of opposite magnetization of different shapes and sizes in the background of the metastable polarized state are true-vacuum bubbles. 
The energy of a domain with area $A$ and perimeter $\partial A$ is determined by the negative bulk contribution $2 h_z A$ and the positive interface term $2J\,\partial A$. 
For specific values of $h_z$ and $J$, certain domains become resonant with the initial metastable state.
Resonances occur when
\begin{equation}\label{eq:res}
2|h_z|A=2J\,\partial A.
\end{equation}
As shown in Fig.~\ref{fig:1}c, this gives one-spin, two-spin, three-spin, and symmetric four-spin resonances at \(h_z=4J\), \(3J\), \(8J/3\), and \(2J\), respectively. Their amplitudes scale parametrically as \(h_x^A\), so the one-spin resonance is the strongest, followed by the two- and three-spin resonances. This hierarchy is directly visible by holding \(h_z\) constant at various magnitudes while ramping $h_x(t)$ from 0 to a small value and back in Fig.~\ref{fig:1}d, where the strongest features correspond to the lowest-order processes, and the snapshots on the right confirm the associated bubble geometries.

The resonance at \(h_z=2J\) is special because it not only allows four-spin nucleation, but also first-order growth of an already existing bubble. For an up-spin adjacent to a down-spin seed, the local Ising contribution from its neighborhood is \(2J\), so the effective longitudinal term becomes \(-\hat{\sigma}_i^z(2J-h_z)\). Geometrically, the move increases the bubble area by one site while creating two additional domain-wall segments, increasing the energy by $4J$. This increase is exactly offset by the longitudinal field at $h_z = 2J$, making the growth process resonant.
Once a seed is present, the bubble can therefore spread through single-spin flips, a first-order process in $h_x$. In comparison, spontaneous nucleation of the four-spin bubble is fourth order, so growth dominates over nucleation by approximately three orders of magnitude in the experimentally relevant regime.

To probe the growth regime directly, we prepare a single flipped spin at the center of the system and ramp the global \(h_z\) through the \(h_z=2J\) resonance while \(h_x\) remains finite, to isolate the fast interface dynamics. The time spent in the resonant regime is estimated to be of order tens of nanoseconds, consistent with the coherent operating regime established in previous annealer-based quantum simulations~\cite{king2022coherent,king2023quantum,king2025beyond}. Further details are given in the SI~\cite{SM}. Representative single-shot configurations are shown in Fig.~\ref{fig:1}e. Starting from the seeded defect, the true-vacuum region expands in a ramified, fractal-like pattern consistent with the local resonance rule: only interface spins in the proper environment can flip resonantly. At later times, weak violations of this ideal pattern appear, signaling the onset of off-resonant processes, but the dynamics remain dominated by the resonant growth channel. Averaging over many realizations, Fig.~\ref{fig:1}f reveals a growing, approximately isotropic domain, centered on the seed. Thus, the \(h_z=2J\) resonance gives rise not merely to isolated resonant moves, but to a robust growth regime with an emergent propagating interface.

\section*{True vacuum bubble growth}

\begin{figure}
    \centering
    \includegraphics[width=\columnwidth]{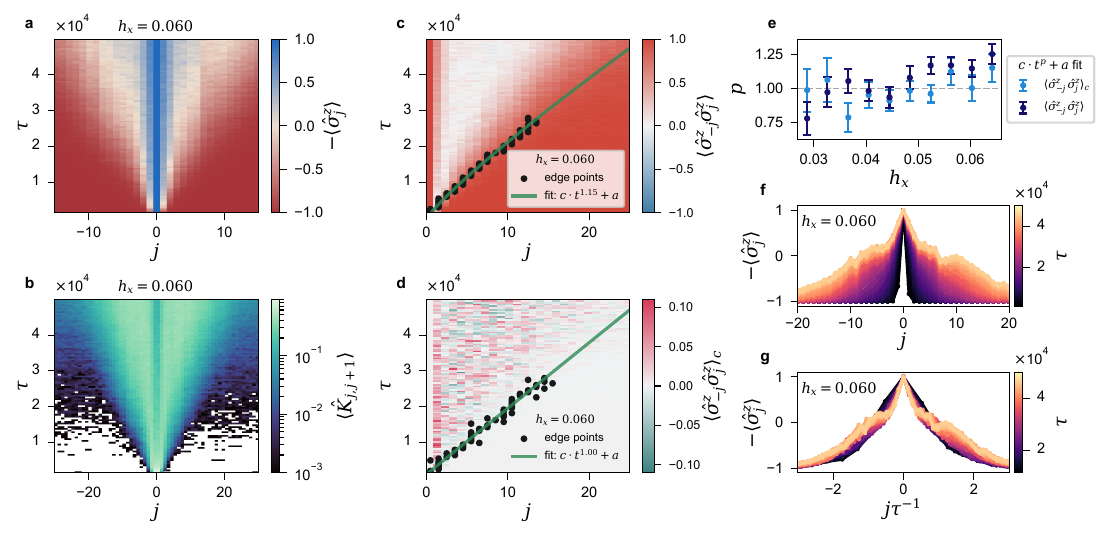}
    
\caption{\textbf{Experimental light-cone dynamics of resonant bubble spreading on the quantum annealer.}
\textbf{a,} Time evolution of the measured local magnetization profile \(-\langle \hat{\sigma}^z_j\rangle\) along a horizontal cut through the center of the two-dimensional system, shown as a function of distance \(j\) from the initially prepared central seed and evolution time \(\tau\), for \(h_x=0.06\). The bright central region corresponds to the expanding true-vacuum bubble, whose front forms a clear light-cone structure.
\textbf{b,} Corresponding nearest-neighbor kink density \(\langle \hat{K}_{j,j+1}\rangle\), showing the propagation of the domain-wall fronts that bound the bubble. The two fronts move approximately symmetrically away from the seeded center and directly visualize the spreading interface.
\textbf{c,} Equal-time correlation function \(\langle \hat{\sigma}^z_{-j}\hat{\sigma}^z_j\rangle\), where \(\hat{\sigma}^z_j\) denotes the single-site magnetization at horizontal distance \(j\) from the central initialized bubble site. The black points mark the extracted edge positions of the correlation front, and the green line shows a power-law fit \(c\cdot\tau^{p}\) with exponent \(p\approx1.15\), indicating nearly ballistic spreading with a weak apparent superlinear correction over the experimentally accessible times.
\textbf{d,} Connected equal-time correlator \(\langle \hat{\sigma}^z_{-j}\hat{\sigma}^z_j\rangle_{\rm c}\). The black points again mark the extracted edge positions, while the green line is a power-law fit \(c\cdot\tau^{p}\) with exponent \(p\approx1.00\). The close agreement with \textbf{c} shows that the front dynamics is robust with respect to disconnected background contributions.
\textbf{e,} Growth exponent \(p\) extracted from power-law fits to the edge positions of \(\langle \hat{\sigma}^z_{-j}\hat{\sigma}^z_j\rangle\) and \(\langle \hat{\sigma}^z_{-j}\hat{\sigma}^z_j\rangle_{\rm c}\) as a function of \(h_x\). The dashed horizontal line marks purely ballistic propagation, \(p=1\). Both correlators yield exponents close to unity across the explored range of transverse fields, consistent with nearly ballistic spreading.
\textbf{f,} Raw magnetization profiles \(-\langle \hat{\sigma}^z_j\rangle\) at different times \(\tau\), showing the outward propagation of the bubble front. The approximately triangular profile in the laboratory frame reflects a front moving with nearly constant velocity while broadening in time.
\textbf{g,} The same data as in \textbf{f}, replotted against the scaled coordinate \(j\tau^{-1}\). The approximate collapse confirms ballistic scaling of the front position, while the residual deviations near the interface encode the slower broadening of the wavefront.
}
    
    \label{fig:2}
\end{figure}

We further characterize the true vacuum bubble growth obtained using the ramp protocol with an artificial initial seed at multiple locations, used for averaging. As shown in Fig.~\ref{fig:2}a, the local magnetization profile, \(\langle \hat{\sigma}^z_j\rangle\), measured along a horizontal cut through the center, develops a clear light-cone structure. The initially localized seed spreads outwards approximately symmetrically, forming an expanding true-vacuum domain inside the false-vacuum background. Figure~\ref{fig:2}b shows the corresponding nearest-neighbor kink density \(\hat{K}_{j,j+1}=(1-\hat{\sigma}^z_j\hat{\sigma}^z_{j+1})/2\). Its sharp expanding front matches the light-cone in the magnetization which demonstrates that the growth is controlled by the motion of a well-defined interface.
For a standard expanding bubble, one would expect the kink density in the bulk to decrease to zero. However, we find a large kink density in the bulk, suggesting a rich internal structure of the growing false vacuum domain, consistent with the ramified pattern.

To characterize the propagation more quantitatively, we study the equal-time correlation function $\langle \hat{\sigma}^z_{-j}\hat{\sigma}^z_j\rangle$ and its connected counterpart $ \langle \hat{\sigma}^z_{-j} \hat{\sigma}^z_j\rangle_{\rm c} = \langle \hat{\sigma}^z_{-j}\hat{\sigma}^z_j\rangle - \langle \hat{\sigma}^z_{-j} \rangle \langle \hat{\sigma}^z_j\rangle$, where \(\hat{\sigma}^z_j\) denotes the single-site magnetization at horizontal distance \(j\) from the central initialized bubble site. The resulting space-time maps are shown in Fig.~\ref{fig:2}c,d. In both cases, the correlation front propagates linearly to a good approximation, allowing us to define an edge position as a function of time. Fitting the extracted front position to a power law \(c\,\tau^p\) gives exponents \(p\approx1.15 \pm 0.06\) for the full correlator and \(p\approx1.00 \pm 0.10\) for the connected correlator. Repeating the same analysis for different values of \(h_x\), Fig.~\ref{fig:2}e shows that the extracted exponent remains close to \(p=1\) across the explored range. The larger deviations at small \(h_x\) arise from slower dynamics, where the corresponding changes in the magnetization profile are harder to resolve and increase the fit uncertainty. By contrast, at large \(h_x\), the fitted exponent is systematically shifted above \(p=1\) because the bubble reaches the system boundary and self-interferes through the periodic boundary conditions. These data indicate that the resonant bubble grows nearly ballistically in the experimentally accessible time window.

This conclusion is further illustrated by the magnetization profiles in Fig.~\ref{fig:2}f. As time increases, the front moves to larger distances while retaining an approximately self-similar shape. We find a collapse by rescaling the coordinate \(j\tau^{-1}\) in Fig.~\ref{fig:2}g, where \(\tau\) is the duration of the linear $h_z(t)$ ramp and is proportional to the time spent in the resonant regime, confirming that the dominant scaling is ballistic. At the same time, the residual deviations from perfect collapse near the interface indicate that the front is not rigid. We address this observation in the next section.

The results presented here are expected to generalize to larger scales and higher dimensions in a weak transverse field regime $h_x\ll J$. In $d=2$ dimensions, similar resonances occur at $\vert h_z\vert=2J/l$, where filaments with thickness $l$ spread ballistically from an initial seed of size $4l^2$ with velocity $h_x^l$. The rate of initial seed nucleation scales with its volume ($h_x^{4l^2}$), while the rate of filament spreading scales with the square root of the volume ($h_x^l$), thereby governing the late-time dynamics. A similar resonant phenomenon is also expected in $d>2$, where the analogous symmetric initial seed would be a (hyper-)cube with side $\frac{d}{d-1}l$ 
and the nucleation rate scaling as $\sim h_x^{\left(\frac{d}{d-1}l\right)^{d}}$.
The corresponding resonances appear at $\vert h_z\vert=2J(d-1)/l$ where filaments with thickness $l$ would grow ballistically with the rate that scales as $\sim h_x^{l^{d-1}}$, still dominating the dynamics in any $d$. The ratio of nucleation and filament growth rate exponents is $l \left(\frac{d}{d-1} \right)^{d}$, implying that filament growth becomes even faster compared to nucleation with increasing dimensionality. 

\begin{figure}
    \centering
    \includegraphics[width=\columnwidth]{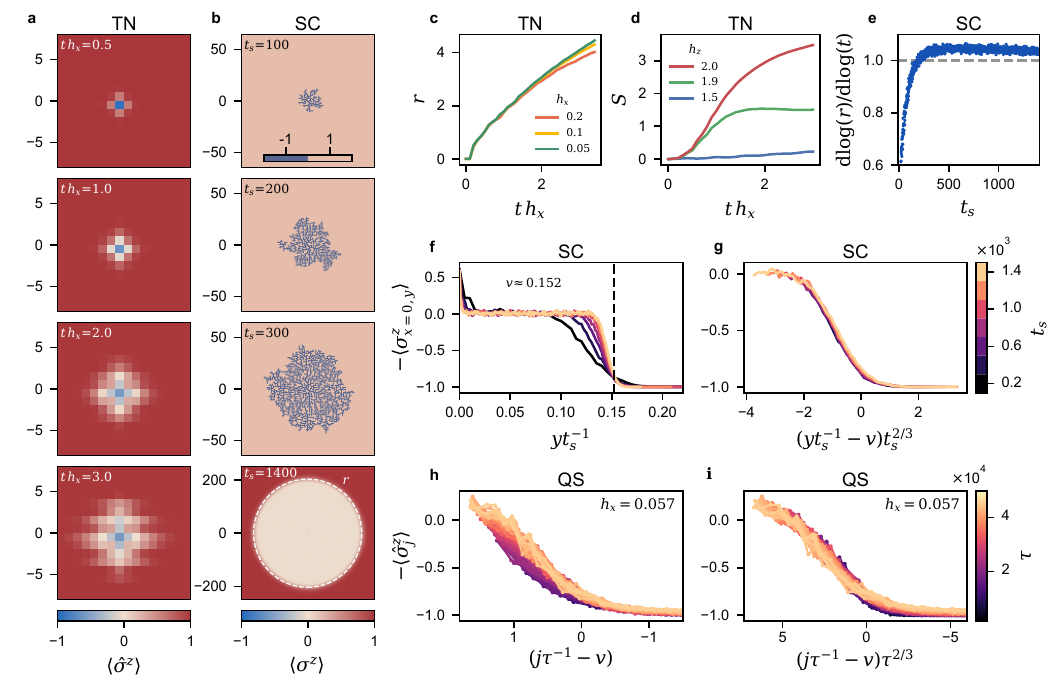}

\caption{\textbf{Ballistic growth and universal interface broadening of resonantly spreading bubbles.}
\textbf{a,} TTN simulations of the magnetization \(\langle \hat{\sigma}^z\rangle\), showing an expanding true-vacuum domain with a broadening front.
\textbf{b,} SC simulations of the effective resonant model showing representative snapshots of the magnetization \(\sigma^z\) at increasing times, with the averaged local magnetization \(\langle \sigma^z_{x,y}\rangle\) at long time shown at the bottom. The dashed circle has radius $r$ equal to $\sqrt{2}$ times the second moment of the bubble profile, demonstrating that the expanding domain remains approximately isotropic. 
\textbf{c,} Bubble radius extracted from the TTN simulations, plotted as \(r\) versus the scaled time \(t h_x\). The data collapse for several values of \(h_x\) confirms that the growth timescale is given by \(t h_x\), as expected for the first-order resonant growth process generated by the effective Hamiltonian.
\textbf{d,} Logarithmic derivative \(d\log(r)/d\log(t)\) extracted from the SC simulations. The effective exponent remains close to unity over the accessible time window and decreases only weakly at late times, consistent with asymptotically ballistic radial growth.
\textbf{e,} Bipartite entanglement entropy \(S\) from the TN simulations as a function of \(t h_x\). The entropy growth remains moderate over the accessible time window, explaining why tensor-network simulations can faithfully capture the early-time quantum dynamics while becoming limited at later times and larger systems. 
\textbf{f,} Collapse of the SC magnetization profile along the line \(x=0\), plotted as \(-\langle \sigma^z_{x=0,y}\rangle\) versus \(y/t\). The good overlap of the profiles at different times demonstrates a well-defined ballistic front propagating with velocity \(v\approx 0.152\), while deviations near the front indicate additional broadening beyond a rigidly translating profile.
\textbf{g,} Scaling collapse of the SC wavefront after subtracting the ballistic drift, plotted as a function of \((y/t-v)t^{2/3}\). The collapse is consistent with front broadening governed by the Kardar--Parisi--Zhang (KPZ) scaling exponent \(2/3\).
\textbf{h,} Experimental quantum-simulator (QS) profiles along a one-dimensional cut through the two-dimensional bubble at several evolution times, plotted in the co-moving coordinate \((j/\tau-v)\). The measured magnetization shows the same ballistic front propagation as in the SC and TN calculations.
\textbf{i,} Collapse of the experimental QS wavefront using the KPZ-scaled coordinate \((j/\tau-v)\tau^{2/3}\). The agreement with the stochastic-circuit scaling supports a universal coarse-grained description of the resonant bubble interface, with nearly ballistic radial growth and sub-ballistic KPZ-type front broadening.}
    
    \label{fig:3}
\end{figure}


\section*{Classical modeling of bubble growth}

We support the quantum simulation results with two complementary classical approaches.
First, we simulate the full quantum model of Eq.~\eqref{eq:fullmodel_rewritten} with tree tensor networks (TTNs), a generalization of matrix product states that better represent long-range correlations in higher-dimensional systems~\cite{Silvi2010}. TTNs allow us to simulate the full quantum dynamics with controlled truncation errors, albeit with limitations on the accessible system sizes and total evolution times~\cite{Haegeman2016, Bauernfeind2020}. 
To access the late-time properties of bubble growth beyond the reach of TTNs, we additionally develop an effective stochastic circuit (SC) model of bubble dynamics. In the SC approach, we initialize a classical configuration with a single lattice site flipped into the true-vacuum state and evolve it in discrete time steps denoted by $t_s$. At each step, any site $j$ is acted on with a constrained spin-flip operator, preserving the resonance condition, with probability $p$ and left unchanged with probability $1-p$. The crucial point is that each update maps a single computational-basis configuration to another computational-basis configuration, and can thus be simulated classically. Further details about the two types of classical simulations can be found in Methods.

To isolate the intrinsic bubble-growth dynamics from the extraneous ramp protocol effects present in quantum simulations, we consider instantaneous quenches from an initial product state in which a single spin is flipped into the true-vacuum state. We then evolve under the full Hamiltonian at finite $h_x$ and fixed $h_z=2J$. The evolution of the magnetization, computed by TTNs, is shown in Fig.~\ref{fig:3}a, while a representative trajectory of the SC simulation is shown in the top three panels of Fig.~\ref{fig:3}b. Each trajectory exhibits the same tree-like pattern reminiscent of the experimental data in Fig.~\ref{fig:1}. 

The bottom panel of Fig.~\ref{fig:3}a reveals the breakdown of TTN calculations: the bubble seemingly grows faster in certain preferred directions. Such anomalous interface effects are a consequence of mapping the 2D system into a tensor network that does not natively represent the lattice geometry~\cite{Bellwood2025}. This is a sign that the simulation becomes unreliable for $t h_x \gtrsim 3$ for this system of $16 \times 16$ spins. On the other hand, the SC dynamics in Fig.~\ref{fig:3}b provides a window to the coarse-grained description of the long-time growth regime, where the dynamics is dominated by the accumulation of many local resonant events and the relevant observables are statistical properties of the growing bubble. This allows us to access the statistical growth dynamics of bubbles in systems containing hundreds of thousands of spins and over thousands of time steps. The magnetization, averaged over many SC realizations, is shown in the bottom panel of Fig.~\ref{fig:3}b. Interestingly, it demonstrates a remarkable spherical invariance despite the underlying $D_4$ symmetry of the square lattice.


The evolution of the bubble size $r$, computed via TTNs from the second moment of $\langle \hat{\sigma}^z\rangle$, is shown in Fig.~\ref{fig:3}c for a few values of $h_x$. This data suggests a ballistic growth with a characteristic timescale of $h_x$, indicating that the growth process is driven by the first-order resonance, rather than by quantum coarsening or other diffusive processes. To extract the long-time features of the growth, we focus on the SC simulation of the same quantity $r$ (see Methods). Figure~\ref{fig:3}d plots the log derivative of $r$ over time, displaying close agreement and convergence towards the ballistic growth exponent of 1. 

Despite the fact that the main features of the bubble growth can be reproduced by SC simulations, the underlying dynamics generate a considerable amount of entanglement and should thus be understood as a genuinely many-body quantum process. Figure~\ref{fig:3}e shows the evolution of the half-cut entanglement entropy $S$ for quenches of $h_z$ to the values indicated in the legend, computed via TTNs. As expected, the state remains approximately frozen when there are no resonant processes available ($h_z=1.5J$)~\cite{balducci2022localization}, while the resonant process at $h_z=2J$ triggers very fast entanglement growth. Interestingly, for a quench close to the resonant point ($h_z=1.9J$), $S$ initially grows at the same rate as in the resonant case, but quickly tapers off and saturates at a smaller value. This might be due to the confinement of excitations, expected at small $h_x$ far from the critical point~\cite{pavesic2025constrained, borla2026microscopic}.

It is inherently difficult to directly compare TTN results with our quantum simulations because the precise $h_z(t)$ and $h_x(t)$ ramps are unknown, even though they have a strong influence on the exact shape of $M(t)$. However, we consistently find a collapse of $M(t)$ with a rescaled time axis $h_x^2t$ in quantum simulation, contrary to the expected $h_xt$ scaling. We therefore performed analogous $h_z(t)$ ramps through the resonance using TTNs and found that the $h_x^2t$ scaling law emerges there as well, see SI~\cite{SM}. This is explained by the quantum nature of the resonance, where perturbation theory predicts a finite resonance width $\delta h_z\propto h_x$. Then the time spent in the resonance during the ramp is $\delta t\propto h_x$, and the scaling law of the dynamics needs to be appropriately modified as $M(t)\sim h_x\delta t\propto h_x^2t$, suggesting agreement between TTN and quantum simulation. 

An alternative way to test for the ballistic  bubble growth is to collapse the data along a single row or column, as is done in Fig~\ref{fig:3}f. Assuming a constant propagation speed leads to a clean collapse everywhere except near the center of the system and around the wavefront. This indicates that the latter broadens more slowly than ballistically. Performing a second collapse of the wavefront data, Fig~\ref{fig:3}g, we find that the best results are obtained for $t_s^{2/3}$. This strongly suggests that our stochastic circuit belongs to the KPZ universality class, which describes transport properties of a variety of quantum systems~\cite{znidaric2011transport,ljubotina2017spin,ilievski2018superdiff}. This class was first developed to describe the growth of interfaces in classical systems, which is indeed similar to the process at play here~\cite{KPZ}.  We perform the same collapses on the quantum simulation data, shown in Fig.~\ref{fig:3}h,i, and find a good agreement both for the ballistic growth of the bubble size and the KPZ coarsening of the interface, see the SI~\cite{SM} for a detailed analysis of the optimal exponent across the entire quantum simulation dataset.

\section*{Discussion and outlook}

We have studied a growth-dominated regime of FVD in two dimensions on a superconducting quantum annealer. At the resonance \(|h_z|=2J\), single-spin flips at the boundary of an existing true-vacuum seed become energy conserving, so growth already proceeds at first order in \(h_x\), while spontaneous nucleation of new bubbles remains parametrically slower. This produces an approximately three orders of magnitude separation between growth and nucleation. In our quantum simulations, the resulting dynamics appear as a clear light-cone structure and nearly ballistic spreading of a seeded bubble, while classical modeling captures the accompanying sub-ballistic interface broadening consistent with KPZ scaling. The essential physics, therefore, lies not in the formation of the initial seed but in its subsequent resonant expansion and fluctuating interface growth described by KPZ universality. 

Several observations indicate that the annealer probes genuinely quantum dynamics. First, the estimated time spent in the dynamical resonant regime is of order a few tens of nanoseconds, consistent with the timescales on which coherent behavior has previously been demonstrated on the same class of annealers~\cite{king2022coherent,king2023quantum,king2025beyond}; see SI for the estimation. 
Second, the observed scaling $M(t)\sim h_x^2t$, seen in both TTN calculations and quantum simulations, arises from resonance broadening, an intrinsically quantum effect.
Third, we find that the classical-energy contribution remains approximately conserved over the experimentally relevant time window, showing that the observed growth stays concentrated near the resonant manifold rather than being dominated by generic off-resonant relaxation due to thermalization.
These points, together with the agreement with approximate classical methods, provide evidence that the annealer realizes the targeted quantum many-body dynamics of the full model over the experimentally relevant time window.
Its ability to simulate large systems is particularly important for extracting the late-time properties of bubble growth, where finite-size effects can otherwise obscure the asymptotic behavior.


At the same time, our results expose an intriguing puzzle. The microscopic dynamics generate substantial entanglement and exhibit signatures consistent with coherent quantum evolution, yet the asymptotic growth laws, i.e., nearly ballistic expansion and KPZ-type interface broadening, are already captured by a SC description of the effective resonant model. This suggests that the large-scale statistical properties of resonant bubble growth may be remarkably insensitive to many microscopic quantum details. Determining precisely how quantum effects influence the asymptotic dynamics, therefore, remains an important open question. Moreover, it is interesting to consider if and how the KPZ universality class changes in higher dimensions $d>2$, where the filaments are growing $(d-1)$-dimensional surfaces. Addressing these questions will likely require new dynamical control capabilities of quantum annealers that are not yet publicly available~\cite{deshpande2026analog}, or significant advances in numerical methods for higher-dimensional quantum dynamics, where current classical approaches remain substantially more limited.


More broadly, the present study should be viewed as a controlled isolation of one particular post-nucleation growth channel, as part of a larger class of similar resonances that appear at other longitudinal field values and in arbitrary dimensions. Generic FVD process will, in general, involve a mixture of nucleation and multiple competing growth channels. This suggests that resonant accretion may provide a broader route to growth-dominated metastable decay whenever local interface moves become energetically favored. An important outlook is therefore to understand whether an analog of this mechanism can arise in quantum field theory, where true-vacuum expansion is usually assumed to be pressure-driven rather than mediated by stochastic accretion. Establishing when such locally enhanced interface growth can compete with conventional bubble-wall motion would connect the resonant mechanism identified here to more generic FVD settings.


\section*{Acknowledgements}
We would like to acknowledge the helpful discussions with Miha Nevemšek and Lorenzo Ubaldi.
J.-Y.D.~acknowledges funding from the European Union’s Horizon 2020 research and innovation programme under the Marie Sk\l odowska-Curie Grant Agreement No.~101034413. Z.P. acknowledges support by the Leverhulme Trust Research Leadership Award RL-2019-015 and EPSRC Grant UKRI3780.
M.L. acknowledges support by the Deutsche Forschungsgemeinschaft (DFG, German Research Foundation) under Germany's Excellence Strategy -- EXC-2111 -- 390814868.
L.P. acknowledges support by the Horizon Europe program HORIZON-CL4-2022-QUANTUM-02-SGA via the project 101113690 (PASQuanS2.1), and the computational resources of the INFN Padova HPC cluster, funded by the NextGenerationEU Terabit project on PNRR – Avviso n. 3264 "per il Rafforzamento e creazione di Infrastrutture di Ricerca", Missione 4, "Istruzione e Ricerca". K.M. acknowledges support from the project Jülich UNified Infrastructure for Quantum computing (JUNIQ) that has received funding from the German Federal Ministry of Education and Research (BMBF) and the Ministry of Culture and Science of the State of North Rhine-Westphalia. G.H. would like to acknowledge the financial support from ARIS, P1-0040 Nonequilibrium Quantum System Dynamics. G.H. and J.V would like to acknowledge the financial support from ERC AdG HIMMS – Hidden metastable mesoscopic states in quantum materials. J.V. would like to acknowledge the financial support from ARIS, P1-0416 Physics of Quantum Technologies and J1-70063 Quantum Simulation of Non-equilibrium Phenomena on Quantum Devices. The authors gratefully acknowledge the Jülich Supercomputing Centre (https://www.fzjuelich.de/ias/jsc) for funding this project by providing computing time on the D-Wave Advantage™ System JUPSI through the Jülich UNified Infrastructure for Quantum computing (JUNIQ). 

\section*{Author Contributions Statement}
G.H. performed and analyzed the quantum simulations. L.P. performed and analyzed classical modeling using tensor networks. J.-Y.D. and M.L. conducted and analyzed classical modeling using stochastic circuits. J.V. conceptualized the study. All authors participated in the discussions of the results and writing of the manuscript.

\section*{Competing Interests Statement}

The authors declare no competing interests. 

\newpage
\cleardoublepage

\bibliography{biblio.bib}

\appendix

\section*{Methods}

\subsection*{Quantum simulation on D-Wave's quantum annealer}

Our quantum simulations utilized the D-Wave Advantage2\_system1.6 and Advantage2\_system1.13 quantum annealers, comprising 4593 and 4579 physical qubits, respectively. Of these, 4058 and 4044 physical qubits were used to implement the logical torus geometry shown in Fig.~\ref{fig:1}a,b. The embedding realizes a square lattice with periodic boundary conditions in both directions, while bypassing missing qubits on the quantum processing unit.

The programmable D-Wave Hamiltonian takes the form~\cite{dwave2020TechnicalDescription}
\begin{equation}
\hat{\cal H}_{\rm DW}(t)= -\frac{A(s)}{2}\sum_i \hat{\sigma}_i^x
+ \frac{B(s)}{2}\left(g(t)\sum_i h_i \hat{\sigma}_i^z + \sum_{i>j} J_{ij}\hat{\sigma}_i^z\hat{\sigma}_j^z\right),
\end{equation}
where \(s=t/t_a\in[0,1]\) is the normalized anneal parameter, \(A(s)\) sets the transverse-field energy scale, \(B(s)\) the problem-Hamiltonian scale, and \(g(t)\) is the time-dependent gain multiplying the linear biases \(h_i\). Restricting \(J_{ij}\) to nearest-neighbor ferromagnetic couplings on the logical lattice, with \(J_{\rm prog}=-0.3\), and taking uniform biases \(h_i=h_{\rm prog}\), the Hamiltonian reduces, after factoring out \(B(s)/2\), to the mixed-field Ising form in Eq.~\eqref{eq:fullmodel_rewritten}, with \(h_x=A(s)/[B(s)J_{\rm prog}]\) and \(h_z(t)=-g(t)h_{\rm prog}/J_{\rm prog}\).

To realize the dynamics, we used a reverse-annealing protocol specified by the schedule \([(0,1),(t_1,s_{h_x}),(t_2,s_{h_x}),(t_3,1)]\), where \(s_{h_x}\) is chosen such that \(h_x=A(s_{h_x})/(B(s_{h_x})J_{\rm prog})\). Starting from a classical spin configuration at \(h_x=0\), the system is reverse annealed to a point with small but finite transverse field, held there for a pause time \(\tau=t_2-t_1\), and then annealed back to \(h_x=0\) for readout. Typical values are \(t_1=1.0\,\mu\mathrm{s}\), \(\tau\) ranging from \(0.25\,\mu\mathrm{s}\) to \(7.0\,\mu\mathrm{s}\), and \(t_3-t_2=200\,\mathrm{ns}\).

We employed two protocols for the longitudinal field using the device’s \(h\)-gain feature. In the first, \(g(t)\) is held constant during the pause, corresponding to a constant \(h_z(t)\); sweeping this value produces the resonance map shown in Fig.~\ref{fig:1}d. In the second, \(g(t)\) is linearly ramped from \(g(t_1)=1\) to \(g(t_2)=-1\), with \(h_{\rm prog}=-0.7\), so that \(h_z(t)\) approaches the \(|h_z|=2J\) resonance while \(h_x\) remains finite. This isolates the fast interface-growth dynamics by ensuring that the system spends only a small fraction of the full reverse anneal near the resonance.

\subsection*{Effective Hamiltonian at the resonance}

At first order in $h_x/J$, the dynamics at the resonance $|h_z|=2J$ is governed by an effective constrained Hamiltonian,
\begin{equation}\label{eq:eff_model}
\hat{H}_\mathrm{eff}=-h_x \sum_{j} \hat{\mathcal{P}}_j\hat{\sigma}^x_j,\quad
\hat{\mathcal{P}}_j=\begin{cases}
    \mathbf{1}, & \text{if } \sum_{i\in \langle i,j\rangle} \sigma^z_i=2,\\
    0, & \text{otherwise},
    \end{cases}
\end{equation}
where the projector \(\hat{\mathcal{P}}_j\) enforces the local resonance condition: a spin may flip only if it has exactly one nearest neighbor in the true vacuum and three in the false vacuum. This projector structure makes the resonance highly selective: isolated nucleation events are strongly suppressed, while growth along an existing interface is allowed. The resulting dynamics is therefore not that of conventional pressure-driven bubble expansion, but a kinetically constrained spreading process in which the FVD  is controlled by local moves at the boundary of preexisting bubbles.

The Hamiltonian~\eqref{eq:eff_model} is obtained using a Schrieffer-Wolf transformation~\cite{Bravyi2011} with $\hat{H}_0$ given by the classical energy $-J\left(\sum_{\langle i,j \rangle}\hat{\sigma}^z_i\hat{\sigma}^z_{j}  -2 \sum_{j} \hat{\sigma}^z_j\right)$ and the perturbation $\hat{V}$ by the dynamical term $- h_x \sum_{j} \hat{\sigma}^x_j$.
The classical energy $\hat{H}_0$ is conserved at all orders in this expansion; however, the bubble growth that we observe is more constrained than simply conserving this quantity. For example, the effective Hamiltonian~\eqref{eq:eff_model} displays Hilbert space fragmentation, and the system only explores the sector of the initial state. 

We are interested in sectors in which there can be multiple initial bubbles, but all of size 1 (i.e., corresponding to a single-site defect). In this case, there are two different emergent conserved quantities: the conservation of the number of bubbles and the absence of loops. The latter means that there are no sequences of neighboring (vertically or horizontally, but not diagonally) down-spins that form a closed path. For example, 
\begin{equation}
    \begin{bmatrix}
        \dab & \dab & \uar & \dab \\
        \dab & \uar & \uar & \dab\\
        \dab & \dab & \dab & \dab
    \end{bmatrix}
    \ \text{and} \
    \begin{bmatrix}
        \dab & \dab & \dab \\
        \dab & \uar & \dab \\
        \dab & \dab & \uar 
    \end{bmatrix}
    \ \text{are not loops, but} \
    \begin{bmatrix}
        \dab & \dab & \dab \\
        \dab & \uar & \dab \\
        \dab & \dab & \dab 
    \end{bmatrix}
    ,\
    \begin{bmatrix}
    \uar & \dab & \dab & \dab &\uar \\
    \dab & \dab & \uar & \dab & \dab \\
    \dab & \uar & \dab & \uar & \dab \\
    \dab & \dab & \dab & \dab & \dab
    \end{bmatrix}
    \ \text{and} \
    \begin{bmatrix}
        \dab & \dab \\
        \dab & \dab
    \end{bmatrix}
    \ \text{are}.
\end{equation}

For any finite value of $h_x$, the higher-order processes are still relevant, even though they are suppressed. 
Notably, already at the second order, the number of bubbles and the number of loops are no longer conserved. For example, the two processes below happen with a rate $-h_x^2/(8J)$. They both create a loop and split a single bubble into two as
\begin{equation}
    \begin{bmatrix}
        \dab & \dab & \dab \\
        \dab & \uar & \uar \\
        \dab & \dab & \uar \\
        \uar & \dab & \dab 
    \end{bmatrix}
    \to 
    \begin{bmatrix}
        \dab & \dab & \dab \\
        \dab & \dab & \uar \\
        \dab & \uar & \uar \\
        \uar & \dab & \dab 
    \end{bmatrix}
\quad \text{and} \quad 
    \begin{bmatrix}
        \dab & \dab & \dab \\
        \dab & \uar & \uar \\
        \dab & \dab & \dab \\
        \uar & \uar & \dab   
    \end{bmatrix}
    \to 
    \begin{bmatrix}
        \dab & \dab & \dab \\
        \dab & \dab & \uar \\
        \dab & \uar & \dab \\
        \uar & \uar & \dab 
    \end{bmatrix}.
\end{equation}
In fact, for the different system sizes and sectors we tested, we find that the combination of the first- and second-order Hamiltonians leads to almost all states with a given value of $\hat{H}_0$ being in the same sector. Thus, for a finite value of $h_x$ used in the experiment, it is expected that configurations violating the constraints appear at later times.
Due to the very strong effect of the second-order term on fragmentation, we can neglect the effect of terms of order 3 and beyond. We note that if we look at a region of false-vacuum (i.e., with only up-spins), the smallest bubbles that can nucleate resonantly are 4-bubbles in a square configuration. As these only occur at fourth order, these processes are highly suppressed and can be neglected if there is any bubble already present.

\subsection*{Tensor network simulations}

In order to fully capture the quantum effects of the bubble growth, we perform simulations with tensor networks.
These provide a controlled approximation to the evolution of the full quantum model.
We use tree tensor networks (TTNs), where the tensors are arranged into a binary tree structure. 
The leaves of the tree contain the physical indices of the lattice sites, whilst the auxiliary tensors in the upper layers help to transfer correlations across distant sites.
This is critical for simulating two-dimensional systems with a one-dimensional tensor network.
In such cases, the lattice has to be mapped down to one dimension using a space-filling curve, which necessarily produces effective long-range interactions.
Nonetheless, TTNs capture long-range correlations better than the standard matrix product states~\cite{Silvi2010}, and it has been recently shown that TTNs can simulate certain dynamical processes of relatively large spin systems in two dimensions~\cite{pavevsic2025scattering, borla2026microscopic}. 

For the simulations of the bubble growth, we start with an initial product state where all spins are aligned down in the $Z$ direction, and a single nucleation center pointing up. 
We evolve the state with the Ising Hamiltonian Eq.~\eqref{eq:fullmodel_rewritten}, with $J=1$, $h_z = 2$ and varying $h_x$. 
This is akin to an instantaneous quench directly into the resonant point. 

We simulate the time evolution with the time-dependent variational principle (TDVP) algorithm~\cite{Haegeman2016, Bauernfeind2020}. 
We use the single-tensor update version of the algorithm, and pad the tensors of the initial product state to the maximal bond dimension.
For TTNs, this is a much more efficient approach compared to the two-site TDVP; because the tensors in the TTN are of rank-3 with all legs at the maximal bond dimension $\chi$, the computational complexity of the single-site TDVP scales with $\mathcal{O}(\chi^4)$, while the two-site TDVP scales as $\mathcal{O}(\chi^6)$.

We have checked that the results are converged in the size of the discretized timestep, and use $dt = 0.01/J$.
All simulations were performed at a set of different bond dimensions, and we checked that the observables of interest are converged with increasing bond dimension. 
Despite the relatively small value of $h_x$, we find that the TTNs cannot accurately capture all features of the dynamics beyond $tJ \sim 5$.
Whilst the average quantities (such as the total magnetization) are well converged, their spatial distributions are not. This is best seen in Fig.~\ref{fig:3}. 
The tensor network simulations were performed using the \textrm{Quantum Tea} package~\cite{qtea_zenodo}. 

\subsection*{Classical stochastic circuit simulations}

To simulate large systems, we turn to a classically simulable stochastic circuit inspired by the effective model (\ref{eq:eff_model}). The basic circuit has a ``brickwork structure'', meaning that for each time step, there are two intermediate steps during which the odd ($x+y$ odd) and even ($x+y$ even) sites are evolved, respectively. This is to ensure that the evolution does not violate the constraint. In each evolution step, all flippable sites (both up to down and down to up) with the correct parity are identified. Each of these is then flipped with a probability $p$ or left unflipped with probability $1-p$. We use $p=0.1$, which keeps the dynamics stochastic and avoids the deterministic limit obtained for $p=1$. We note that the results are insensitive to the exact value of $p$ as long as $p\ll 1$. The only influence of $p$ is a rescaling of time. We show this invariance with respect to $p$ in the SI.

In order to make sure that what we observe is a generic feature of the constraint and not an artifact of our exact evolution scheme, we have also tested an alternative circuit. Instead of having a brickwork structure, each step is composed of a single evolution step in which both odd and even sites can be updated. We consider flips on all sites satisfying the constraint (still with probability $p$) and subsequently resolve ``conflicts'', where flips on neighboring sites would lead to violations of the constraints. We have explored different ways of resolving these ``conflicts'' while maximizing the number of flips, and found that the alternate circuit also shows ballistic growth (with a slightly different velocity) in all cases. See SI for more details and results.

\subsection*{Radially averaged bubble size}

In Fig.~\ref{fig:3} we show the radially averaged bubble size $r$, extracted from TN and SC simulations.
In both cases, it is computed from the local magnetization as
\begin{equation}
    r=\sqrt{2\sum_{x,y} \langle n_{x,y}\rangle\left[(x-x_0)^2+(y-y_0)^2\right]}, \ \text{with} \ x_0=\sum \langle n_{x,y}\rangle x, \ y_0=\sum \langle n_{x,y}\rangle y, 
\end{equation}
where $n_{x,y}=(1-\sigma^z_{x,y})/2$ is equal to one if a site is in the true vacuum (and so part of the bubble) and to zero if it is in the false vacuum. 

\end{document}


\onecolumngrid
\begin{center}
\textbf{\large Supplemental Material for ``Resonant false vacuum decay in two dimensions on a 4000-qubit quantum annealer" }\\[5pt]
\end{center}
\setcounter{equation}{0}
\setcounter{figure}{0}
\setcounter{table}{0}
\setcounter{page}{1}
\setcounter{section}{0}
\makeatletter
\renewcommand{\theequation}{S\arabic{equation}}
\renewcommand{\thefigure}{S\arabic{figure}}
\renewcommand{\thesection}{S\arabic{section}}
\renewcommand{\thepage}{\arabic{page}}
\renewcommand{\thetable}{S\arabic{table}}

\vspace{0cm}

\tableofcontents

\section{QPU embeddings}

\begin{figure}
    \centering
    \includegraphics[width=\columnwidth]{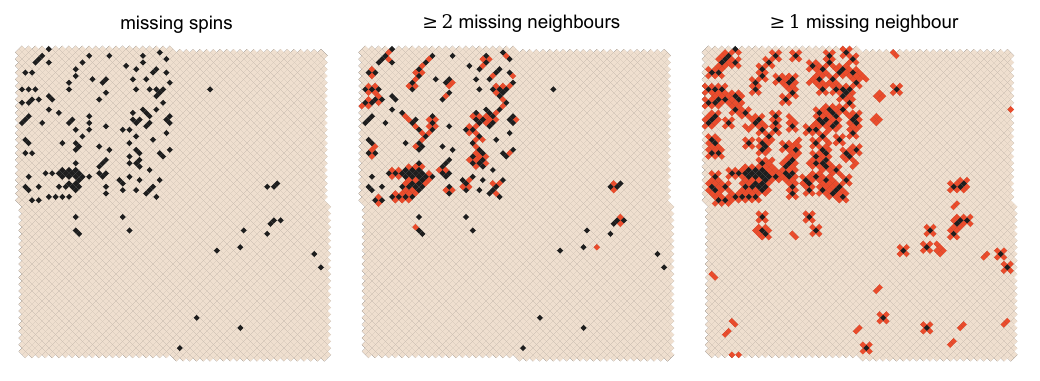}
\caption{Logical square-lattice configurations with periodic boundary conditions realized on the Advantage2\_system1.6 quantum annealer. (left) Black squares indicate missing logical spins arising from unavailable physical qubits. (middle) Orange squares denote logical spins with two or more missing neighboring logical connections due to non-functional physical couplers. (right) Orange squares denote logical spins with at least one missing connection to a neighboring spin resulting from non-functional physical couplers.} 
    \label{fig:1}
\end{figure}

\begin{figure}
    \centering
    \includegraphics[width=\columnwidth]{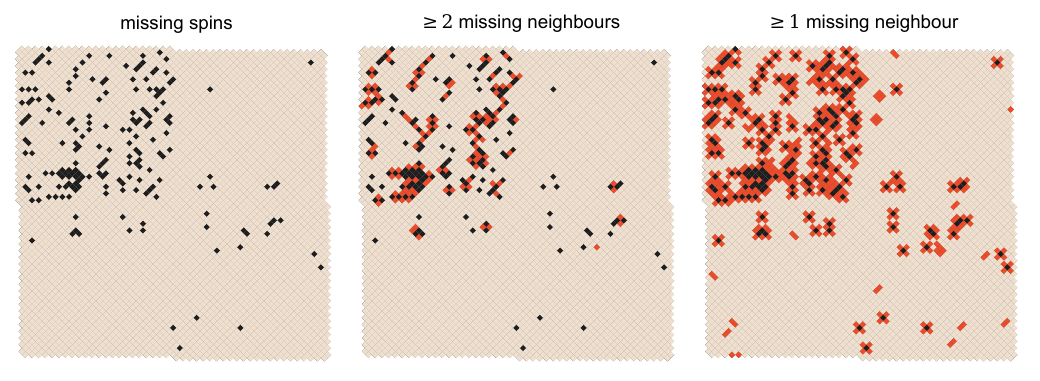}
\caption{Logical square-lattice configurations with periodic boundary conditions realised on the Advantage2\_system1.13 quantum annealer. (left) Black squares indicate missing logical spins arising from unavailable physical qubits. (middle) Orange squares denote logical spins with two or more missing neighboring logical connections due to non-functional physical couplers. (right) Orange squares denote logical spins with at least one missing connection to a neighboring spin resulting from non-functional physical couplers.}
    \label{fig:2}
\end{figure}

We generate an embedding of a two-dimensional torus on 4593 qubits of the Advantage2\_system1.6 graph (Fig.~\ref{fig:1}) and 4579 qubits of the Advantage2\_system1.13 graph (Fig.~\ref{fig:2}) by connecting individual layers of a two-dimensional square lattice. A single layer is constructed by connecting two vertical and two horizontal qubits positioned opposite one another within a single Zephyr cell and then repeating this pattern across all cells. The Zephyr graph architecture allows four independent layers to be constructed using the same Zephyr cells without interlayer couplings \cite{dwave2021Zephyr}. Periodic boundary conditions are implemented by connecting the edges of these layers such that the resulting logical system forms a twisted torus.  

Some qubits and couplers in the full Zephyr graph are absent on the physical device due to fabrication defects. To minimize the impact of these defects on the simulations, and to create larger regions suitable for studying bubble expansion, the connectivity of qubits within defective cells is adjusted so that defects are clustered as much as possible within a single layer. As a result, the corresponding defects are localized within a restricted region of the logical system, allowing larger defect-free regions in the remaining parts of the lattice.  

The Advantage2\_system1.6 was used to perform simulations at constant \(h_z\) while sweeping over different values of \(h_z\). During these simulations, qubits missing more than one coupling to neighboring spins (shown on the right panel of Fig.~\ref{fig:1}), were manually biased with a local field \(h_z\) using the flux bias offset parameter of the D-Wave device such that they did not participate in the dynamics. This was done to preserve the defect-free resonance conditions, since missing spins and couplers introduce additional effects. For example, in a defect-free system, a bubble of size \(1\) nucleates at \(h_z = 4J\), corresponding to the point at which the energy cost of creating four domain walls equals the energy gain from flipping a spin. In the presence of defects, however, a size-\(1\) bubble can nucleate adjacent to a defect already at \(h_z = 3J\), since only three domain walls must be created. Similar effects occur for bubbles of larger sizes.  

The Advantage2\_system1.13 was used to perform simulations studying bubble expansion under a ramp protocol. In these simulations, qubits with two or more missing neighbouring couplings (shown on the middle panel of Fig. Fig.~\ref{fig:2}) were excluded from the dynamics using local \(h_z\) fields in order to isolate the spreading of the manually seeded bubble. Since the dynamics were studied at the resonant condition \(h_z = 2J\), qubits with two missing couplings can themselves become resonant nucleation sites, as flipping such a spin requires the creation of only two domain walls. These flipped spins can then act as seeds for additional bubbles that spread through the resonance mechanism. 

\section{Measurement procedure}
The programmable D-Wave Hamiltonian is given by
\begin{equation}
\hat{\mathcal H}_{\rm DW}(t)=
-\frac{A(s)}{2}\sum_i \hat{\sigma}_i^x
+\frac{B(s)}{2}
\left(
g(t)\sum_i h_i \hat{\sigma}_i^z
+
\sum_{i>j} J_{ij}\hat{\sigma}_i^z\hat{\sigma}_j^z
\right),
\end{equation}
where \(\hat{\sigma}_i^\alpha\) are Pauli operators acting on qubit \(i\), \(h_i\) are programmable longitudinal fields, and \(J_{ij}\) are programmable couplings between qubits \(i\) and \(j\). The couplings are nonzero only for qubits that are physically connected in the quantum processing unit. The parameter \(s=t/t_a\in[0,1]\) is the normalized annealing coordinate, with \(A(s)\) and \(B(s)\) setting the energy scales of the transverse-field and Ising terms, respectively. The annealing schedule \(s(t)\) is specified by a series of user-defined points \((t_i,s_i)\), between which the schedule is linearly interpolated. Similarly, the time-dependent gain function \(g(t)\), defined by a set of control points \((t_i,g_i)\), provides control of the longitudinal fields through the combination \(g(t)h_i\).
To implement the mixed-field Ising model studied here, we restrict \(J_{ij}\) to nearest-neighbor ferromagnetic couplings on the logical lattice and choose a uniform coupling strength \(J_{ij} = J_{\rm prog}=-0.3\). The longitudinal fields are set to be uniform, \(h_i=h_{\rm prog}\). Factoring out the overall scale \(B(s)/2\), the Hamiltonian can then be mapped onto the mixed-field Ising form, with the effective transverse and longitudinal fields given by \( h_x=\frac{A(s)}{B(s)|J_{\rm prog}|}\), \( h_z(t)=-g(t)h_{\rm prog}/J_{\rm prog}\).
This mapping allows both \(h_x\) and \(h_z\) to be controlled independently during the annealing process through the annealing schedule \(s(t)\) and the gain schedule \(g(t)\), respectively. 

To realize the dynamics, we employed a reverse-annealing protocol specified by the schedule \([(0,1),(t_1,s_{h_x}),(t_2,s_{h_x}),(t_3,1)]\), where \(s_{h_x}\) is chosen such that \(h_x=\frac{A(s_{h_x})}{B(s_{h_x})\lvert J_{\rm prog}\rvert}\). Starting from a classical spin configuration at \(h_x=0\), the system is reverse annealed to a point with a small but finite transverse field, held there for a pause time \(\tau=t_2-t_1\), and subsequently annealed back to \(h_x=0\) for readout. We read out the states of all the qubits in the computational basis. Typical values are \(t_1=1.0\,\mu\mathrm{s}\), \(\tau\in[0.25, 7.0]\,\mu\mathrm{s}\), and \(t_3-t_2=200\,\mathrm{ns}\). 

\begin{figure}
    \centering
    \includegraphics[width=0.6\columnwidth]{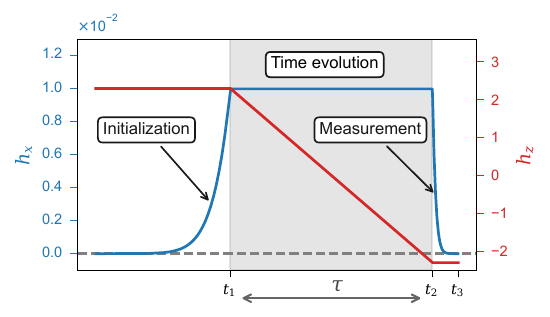}
\caption{Ramp protocol used to characterize bubble expansion dynamics. A configuration containing a single initial seed spin is prepared during a time interval \(t_1 = 1~\mu\mathrm{s}\) by slowly ramping \(h_x\) from zero to a small target value while maintaining a constant positive \(h_z\). Subsequently, \(h_z\) is ramped to the opposite sign over a timescale \(\tau\). Finally, \(h_x\) is reduced to zero as rapidly as possible during the interval \(t_3 - t_2\), after which the final spin configuration is measured. The protocol is repeated \(100\) times for each value of \(\tau\) and for each of the \(10\) initial seed-spin locations, resulting in a total of \(1000\) measurements for each data point.} 
    \label{fig:4}
\end{figure}

Two different protocols for the longitudinal field were implemented using the device's \(h\)-gain feature. In the first protocol, \(g(t)\) is held constant throughout the pause, corresponding to a fixed longitudinal field \(h_z\). We repeat this procedure $100$ times and average over repetitions and individual qubit measurements to get the average magnetization. Repeating the experiment for different values of \(h_z\) allows us to map out the resonance structure shown in Fig.~\ref{fig:5}. In the second protocol, \(g(t)\) is linearly ramped from \(g(t_1)=1\) to \(g(t_2)=-1\), with \(h_{\rm prog}=-0.7\), causing the longitudinal field to approach into the \(|h_z|=2J\) resonance while maintaining a finite transverse field. An example of the schedule is shown in Fig.~\ref{fig:4}. By varying the duration of the ramp, we control the time spent within the resonant region. This allows us to directly probe the resonant interface-growth dynamics and determine how the growth depends on the time near the resonance. The maximum duration of the ramp for all $h_x$ values is set to $5 \times10^4$ qubit time units which are determined by the energy scale as $2 \hbar/(B(s_{h_x})\lvert J_{\rm prog}\rvert)$.  

\begin{figure}
    \centering
    \includegraphics[width=0.33\columnwidth]{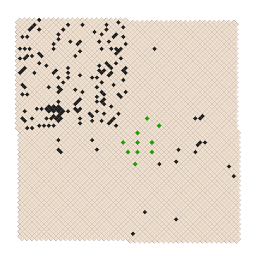}
    \caption{Configuration showing the spin locations used as the initial bubble seeds in the simulations characterizing bubble expansion. Black squares indicate spins absent from the system due to defects, while green squares denote the positions of the initial seeded bubbles.}
    \label{fig:3}
\end{figure}

To study the spreading dynamics in detail, we manually seed the initial bubble as a single flipped spin, which is fixed in place throughout the simulation using a local \(h_z\) field implemented via the flux-bias control parameter. This procedure ensures that the initial position of the bubble is well-defined and identical across realizations, allowing us to systematically study the subsequent time evolution of a single nucleated defect.
In practice, small inhomogeneities between qubits can lead to biases toward particular bubble configurations. To mitigate these effects, we repeat the protocol using $10$ different seed qubits as initial flipped spins. For each seed location, we perform $100$ runs at each ramp duration. The resulting measurements are then averaged over all seed positions, thereby reducing qubit to qubit variations and improving statistical robustness. The locations of all seed spins used in the experiments are shown in Fig.~\ref{fig:3}.

\section{Estimating the time duration of the resonant evolution}
\begin{figure}
    \centering
    \includegraphics[width=\columnwidth]{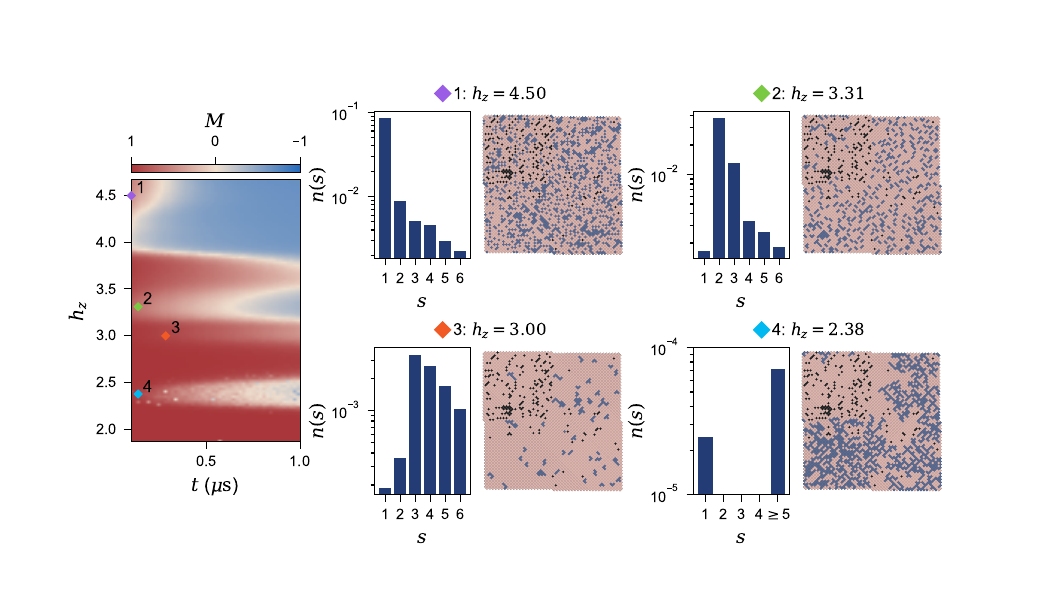}
\caption{(Left) Experimental sweep of the longitudinal field on the annealer, showing the magnetization \(M\) as a function of \(h_z\) and evolution time \(t\) for the constant-\(h_z(t)\) reverse-annealing protocol. (Right)  Representative spin configurations and corresponding bubble-size distributions \(n(s)\) measured at selected resonance points. }
    \label{fig:5}
\end{figure}

To estimate the time spent by the system within the resonance regime, we first determine the resonance width at \(h_x=0.1J\) from the sweep shown in Fig.~\ref{fig:5}. The ramp protocol is designed to terminate at a final value of \(h_z \approx  2.33J\), which approximately corresponds to the center of the resonance peak observed in the sweep. Although the longitudinal field remains near its resonant value during the final measurement quench of \(h_x\), the transverse field is rapidly reduced, strongly suppressing the dynamics and effectively freezing the state of the system. We therefore neglect this contribution when estimating the time spent in resonance. A further complication is the dependence of the resonance width on \(h_x\). However, since the width is estimated from measurements performed at the largest transverse field considered in the simulations, we can treat it as constant, which provides an upper bound on the resonance width and, consequently, on the time spent within the resonant region. The relevant timescale is therefore determined primarily by the duration of the \(h_z\) ramp. For ramp times ranging from \(0.26\,\mu\mathrm{s}\) to \(7\,\mu\mathrm{s}\), and noting that the protocol sweeps the field from \(+h_z\) to \(-h_z\), a simple algebra estimate yields a time within the resonance window of approximately \(8\,\mathrm{ns}\) to \(200\,\mathrm{ns}\) from the shortest ramp length to the longest. 

\section{Longitudinal field sweep}
The sequence of the various resonances is directly visible in the constant-\(h_z\) sweep shown in Fig.~\ref{fig:5}, where the measured magnetization \(M\) is plotted as a function of longitudinal field and evolution time. A detailed scan over \(h_z\) reveals the expected sequence of resonant features, with the lowest-order resonances appearing most prominently and the higher-order two- and three-spin resonances progressively weaker, consistent with their smaller tunneling amplitudes. While the observed resonance structure agrees qualitatively with theoretical predictions, the resonances occur at slightly shifted values of \(h_z\). This discrepancy is most likely due to small differences between the nominal programmed field and the actual longitudinal field realized in the physical device. The effect is further amplified by the use of a slightly reduced programmed coupling strength \(J_{\mathrm{prog}}\), which increases the sensitivity of the resonance positions to such calibration offsets.
To verify that the observed dynamics correspond to the predicted resonant bubble nucleation processes, we analyze both representative spin configurations and the statistics of domain sizes at the resonance points. Specifically, we consider the number density \(n(s)\) of connected domains (bubbles) of the true-vacuum phase containing \(s\) spins, defined as \( n(s)=\frac{N(s)}{N_{\mathrm{sys}}N_{\mathrm{reps}}}\), where \(N(s)\) is the number of bubbles of size \(s\), \(N_{\mathrm{sys}}\) is the total number of spins in the system, and \(N_{\mathrm{reps}}\) is the number of snapshots included in the analysis. Representative snapshots together with the corresponding bubble-size distributions are shown on the right-hand side of the figure. These measurements confirm the expected resonance dynamics: bubbles of the resonant size dominate the distribution at each resonance, while larger bubbles appear with progressively lower density as a consequence of decoherence-driven growth following nucleation. The one-spin resonance is intentionally measured at a slightly offset value of \(h_z\), since its dynamics are fast enough that, even at the shortest accessible evolution times, the system has already largely decayed to the true-vacuum ground state through a combination of single-spin bubble nucleation and subsequent noise-assisted relaxation. This effect is particularly pronounced because the sweep was performed at \(h_x=0.1J\), a value chosen to make even the higher-order resonances experimentally observable. The snapshots further confirm the nature of the individual resonant processes, showing isolated one-spin nucleation near \(|h_z|=4J\), small two- and three-spin bubbles at lower fields, and the onset of the \(2J\) resonance where bubble growth becomes energetically allowed. Taken together, the magnetization data, real-space spin configurations, and bubble-size statistics provide direct experimental evidence that the annealer resolves the predicted sequence of resonant bubble nucleation and growth processes.

\section{Magnetization and classical energy scaling}
\begin{figure}
    \centering
    \includegraphics[width=0.8\columnwidth]{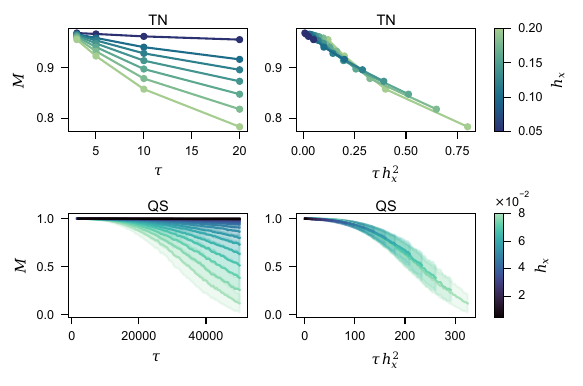}
    \caption{Magnetization measured after the ramp protocol in tensor networks numerical simulations (TN) (top) and in quantum simulations (QS) (bottom) as function of ramp duration $\tau$ (left) and as a function a rescaled time axis $h_x^2 \tau$ (right), plotted for different $h_x$ values. The rescaled curves for both methods collapse onto a single curve.}
    \label{fig:6}
\end{figure}

The observed dependence of the dynamics on \(h_x\) reveals a nontrivial scaling behavior that can be understood from the underlying quantum dynamics. For a system held at a fixed resonant field, the leading-order growth process is governed by an effective Hamiltonian whose dynamics depend on the combination \(h_x t\). In our protocol, however, the seeded bubble evolves while the longitudinal field \(h_z(t)\) is swept through the resonance. In this case, both the experimental data and numerical simulations indicate an effective scaling with \(h_x^2 t\) as shown in Fig.~\ref{fig:6}. This behavior arises because the width of the resonance is itself proportional to \(h_x\). During a linear sweep of \(h_z\), the duration for which the system remains within the resonant region therefore scales as \(h_x t\). Since the resonant growth rate is also proportional to \(h_x\), the total amount of growth accumulated during the sweep is expected to depend on the product of these two factors, leading to an overall scaling in the variable \(h_x^2 t\). Fig.~\ref{fig:6} compares the experimental results with tree-tensor-network simulations performed using a comparable ramp protocol. In the numerical calculations, the ramp is chosen such that it terminates exactly at the resonance, closely matching the experimental procedure. The magnetization is shown both as a function of the ramp duration \(\tau\) and as a function of the rescaled variable \(h_x^2\tau\). While the data obtained at different values of \(h_x\) exhibit distinct behavior when plotted against \(\tau\), they collapse onto a common curve when plotted against \(h_x^2\tau\). The same collapse is observed in the tensor-network simulations, indicating that the dynamics in both cases are controlled by the scaling variable \(h_x^2\tau\).

\begin{figure}
    \centering
    \includegraphics[width=0.8\columnwidth]{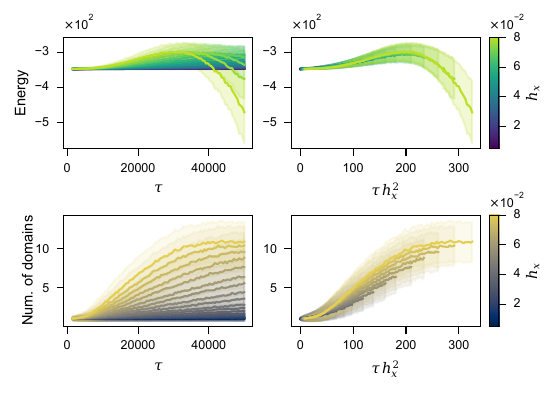}
\caption{Classical energy (top) and number of domains of the true vacuum state (bottom) as a function of $\tau$, plotted for several $h_x$ values (left) and as a function of a rescaled time axis $h_x^2 \tau$ (right). Rescaled energy curves collapse onto a single trajectory. Contrastingly, domain number curves exhibit slower dynamics.}
    \label{fig:7}
\end{figure}

To further probe the coherence of the dynamics, we analyze two quantities extracted from the final spin configurations. The first is the classical Ising energy \(E = h_z \sum_i s_i^z - \sum_{\langle ij \rangle} s_i^z s_j^z, \) where \(s_i=\pm1\) denotes the measured state of spin \(i\). The second is the number of disconnected domains of the true-vacuum phase, obtained by counting all connected clusters of flipped spins. Both quantities are averaged over all snapshots and over the ten different seed-spin locations used in the experiment. The resulting data are shown in Fig.~\ref{fig:7}. At short times, the classical energy remains approximately conserved, indicating that the dynamics remains mostly driven by the resonance condition rather than being dominated by generic off-resonant relaxation processes. This behavior is consistent with coherent resonant growth and provides evidence that the annealer is accessing the intended quantum dynamics. At longer times, deviations from energy conservation become apparent. Initially, these manifest as an increase in the measured classical energy, which coincides with an increase in the number of domains. This behavior can be attributed to noise-induced spin flips that create additional domain walls and hence raise the classical energy. The resulting small bubbles then act as new seed domains, which can themselves undergo resonant expansion. At still later times, the classical energy begins to decrease while the number of domains remains approximately constant. This behavior is consistent with incoherent relaxation toward the true-vacuum ground state, as decoherence progressively fills in the bulk of the already expanded bubbles. A comparison of the scaling behavior of these observables reveals a further connection to the resonant growth mechanism. The classical energy exhibits the same \(h_x^2 t\) scaling observed in the magnetization dynamics, as evidenced by the collapse of energy curves obtained at different values of \(h_x\) onto a single universal trajectory when plotted against \(h_x^2 t\). In contrast, the number of domains displays a somewhat slower scaling behavior, suggesting that domain nucleation and domain growth are governed by partially distinct processes. 

\section{Extended data for characterization of bubble expansion dynamics}
\begin{figure}
    \centering
    \includegraphics[width=\columnwidth]{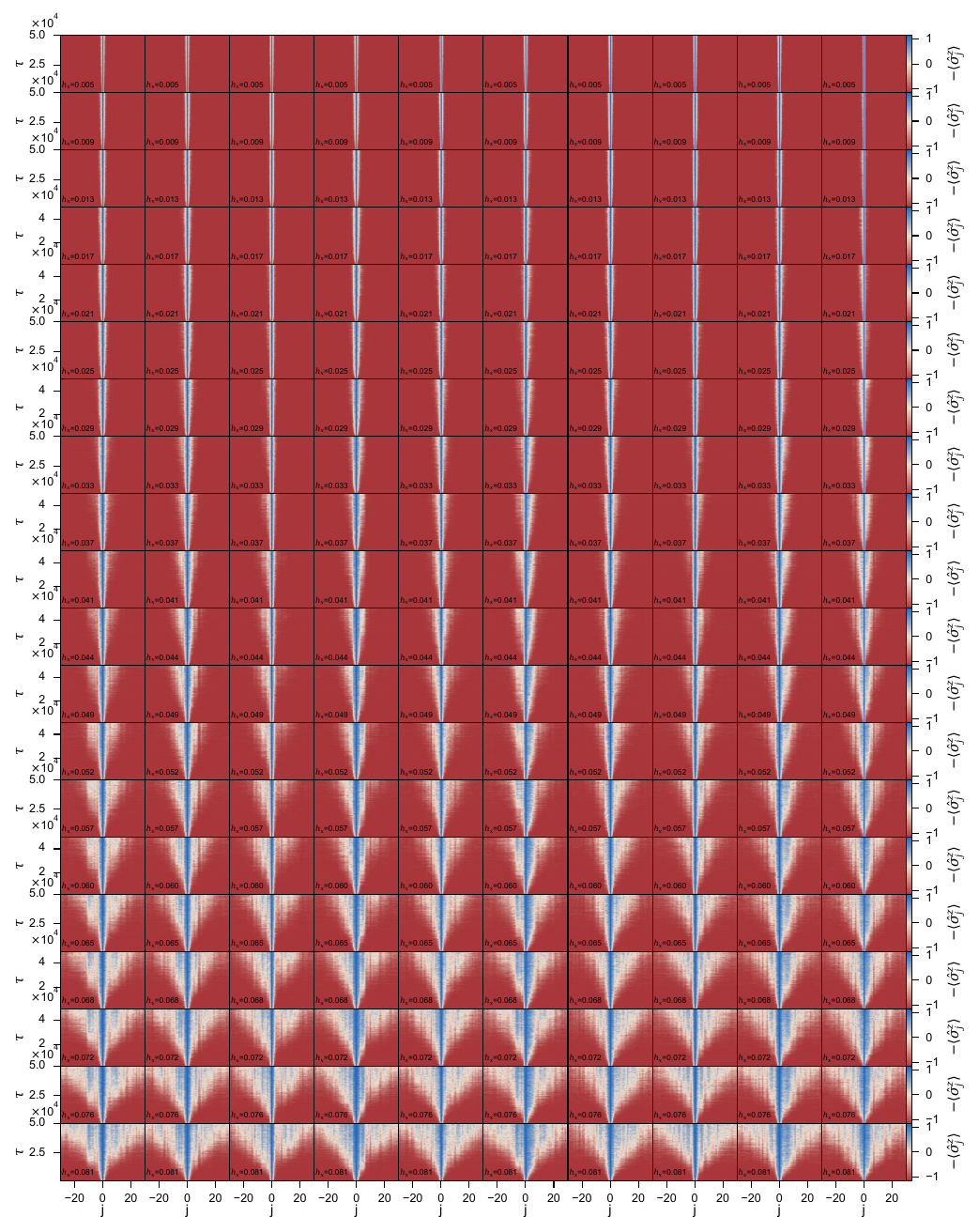}
\caption{Time evolution of the measured local magnetization profile \(-\langle \sigma^z_j\rangle\) along a horizontal cut through the center of the two-dimensional system, shown as a function of distance \(j\) from the initially prepared central seed and evolution time \(\tau\), for all inital seed locations and all measured \(h_x\) values.}
    \label{fig:8}
\end{figure}

\begin{figure}
    \centering
    \includegraphics[width=\columnwidth]{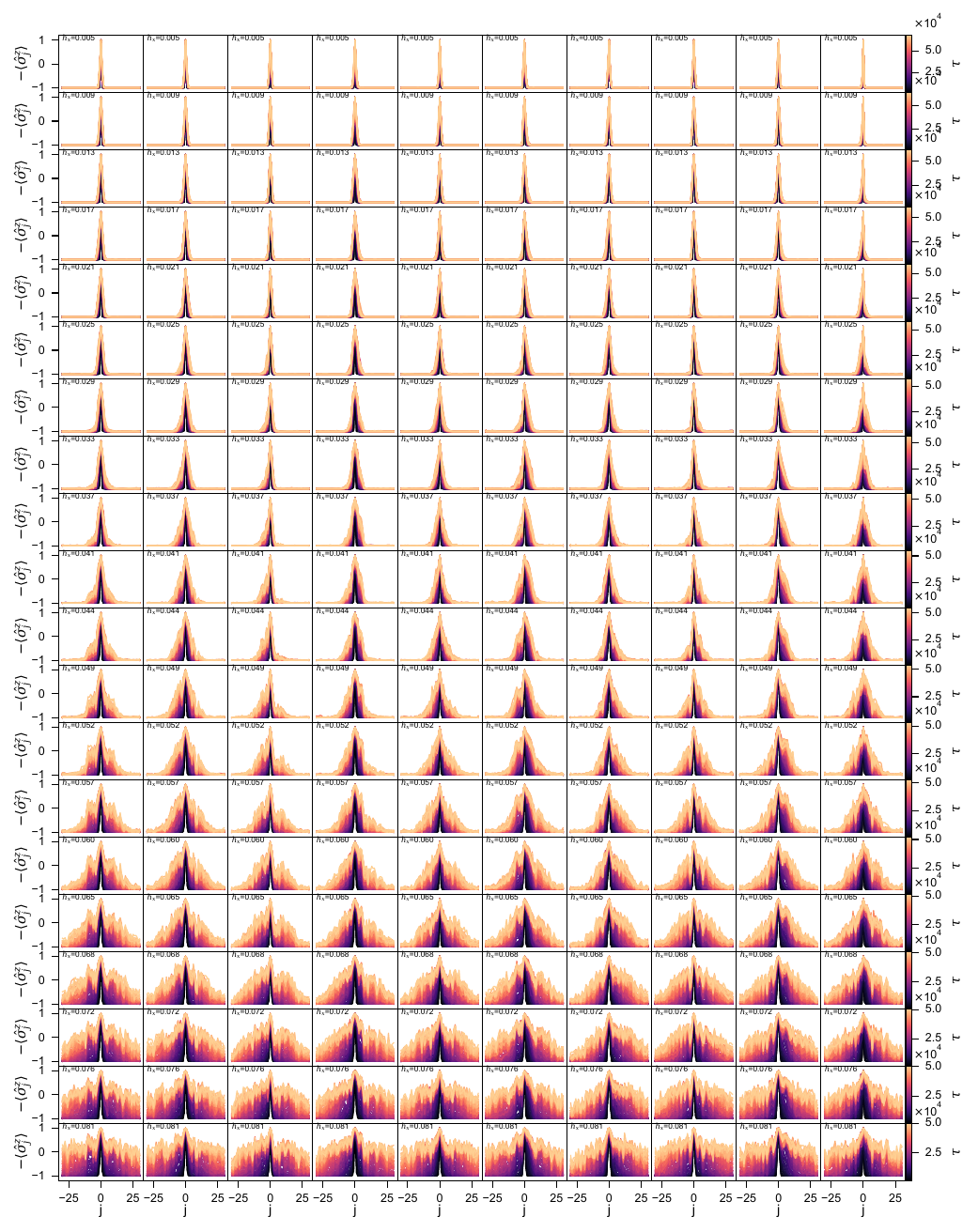}
\caption{Magnetization profiles \(-\langle \hat{\sigma}^z_j\rangle\) at different times \(\tau\) along a horizontal cut through the center of the two-dimensional system, shown as a function of distance \(j\) from the initially prepared central seed and evolution time \(\tau\), for all inital seed locations and all measured \(h_x\) values.}
    \label{fig:9}
\end{figure}

\begin{figure}
    \centering
    \includegraphics[width=\columnwidth]{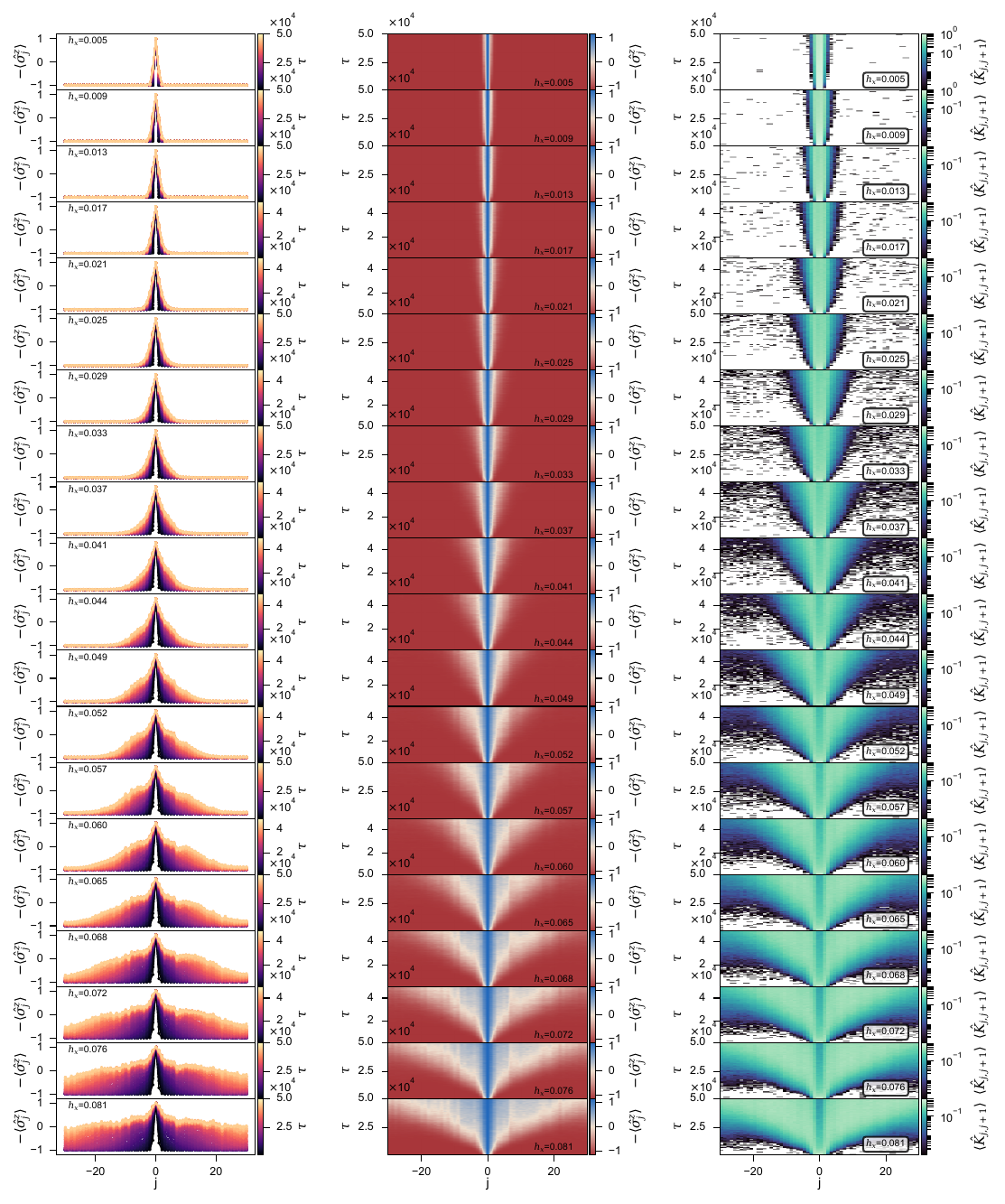}
\caption{Raw magnetization profiles \(-\langle \hat{\sigma}^z_j\rangle\) (left), its time evolution (middle) and nearest-neighbor kink density \(\langle \hat{K}_{j,j+1}\rangle\) (right) along a horizontal cut through the center of the two-dimensional system, shown as a function of distance \(j\) from the initially prepared central seed and evolution time \(\tau\) averaged over all initial seed locations and shown for all measured \(h_x\) values.}
    \label{fig:10}
\end{figure}

The complete dataset for the magnetization, presented both as two-dimensional arrays and as horizontal line cuts through the center of the system, is presented for each initial seed-spin location in Figs.~\ref{fig:8} and \ref{fig:9}. To reduce the effects of local inhomogeneities and defects, the data from all seed-spin locations are combined. The resulting averaged data are presented in Fig.~\ref{fig:10} together with the corresponding kink-density calculations, defined as \(\hat{K}_{i,i+1} = \frac{1}{2}\sum_i \left(1-\hat{\sigma}^z_i\hat{\sigma}^z_{i+1}\right)\).
To characterize the propagation dynamics more quantitatively, we evaluate the equal-time correlation function
\[
\langle \hat{\sigma}^z_{-j}\hat{\sigma}^z_j\rangle
\]
and the connected correlation function
\[
\langle \hat{\sigma}^z_{-j}\hat{\sigma}^z_j\rangle_{\mathrm{c}}
=
\langle \hat{\sigma}^z_{-j}\hat{\sigma}^z_j\rangle
-
\langle \hat{\sigma}^z_{-j}\rangle
\langle \hat{\sigma}^z_j\rangle,
\]
where \(\hat{\sigma}^z_j\) denotes the magnetization of a spin located a horizontal distance \(j\) from the center of the initially seeded bubble.

To extract the position of the bubble boundary, threshold values are chosen for both correlation functions. The boundary position is then defined by the distance \(j\) and evolution time \(\tau\) at which the corresponding threshold is reached. The thresholds used in the analysis are \(0.93\) for the equal-time correlation function and \(5.0\times10^{-5}\) for the connected correlator.

\begin{figure}
    \centering
    \includegraphics[width=\columnwidth]{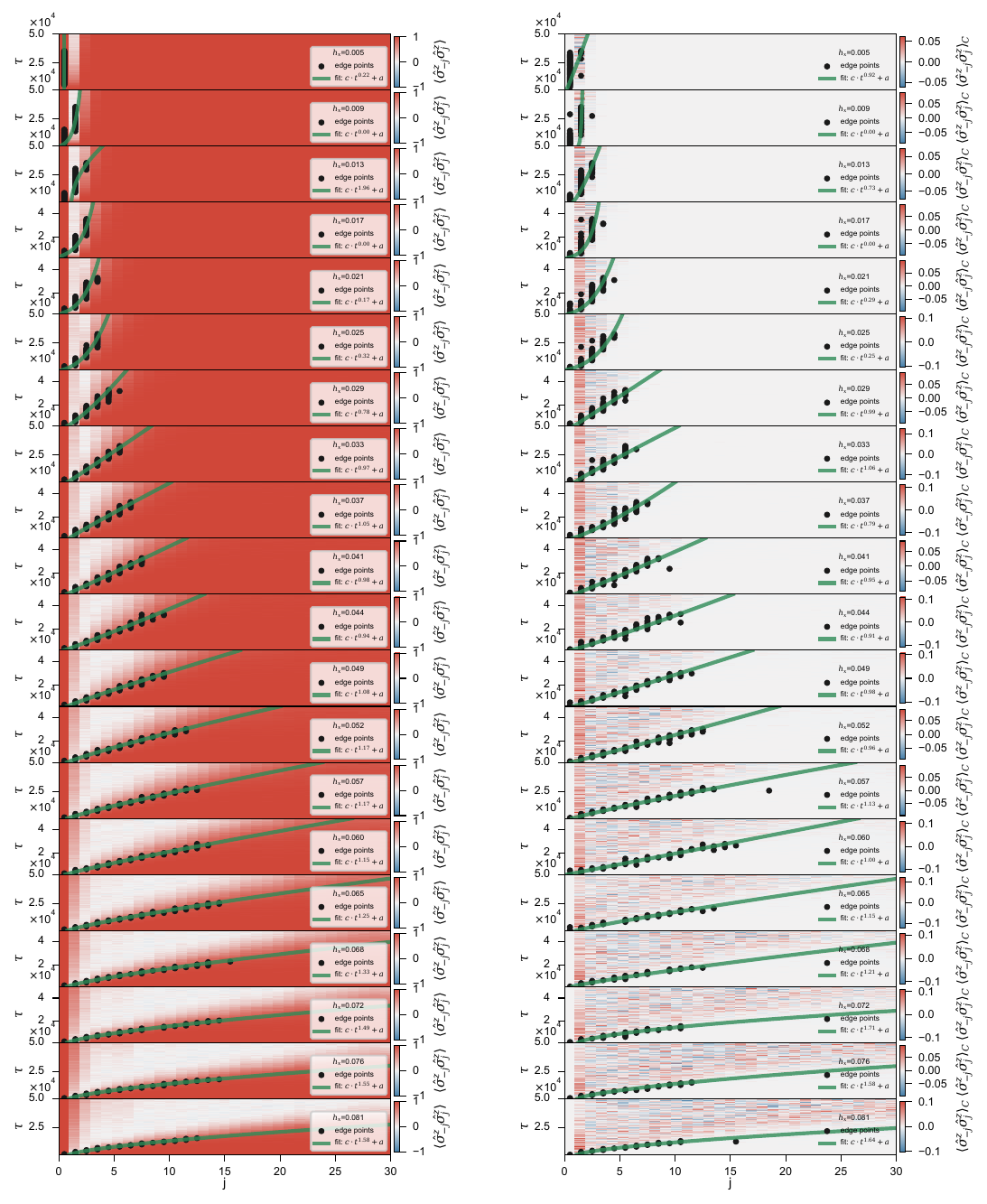}
\caption{Equal-time correlation function \(\langle \hat{\sigma}^z_{-j}\hat{\sigma}^z_j\rangle\) (left) and connected equal-time correlator \(\langle \hat{\sigma}^z_{-j}\hat{\sigma}^z_j\rangle_{\rm c}\) (right), where \(\hat{\sigma}^z_j\) denotes the single-site magnetization at horizontal distance \(j\) from the central initialized bubble site. The black points mark the extracted edge positions of the correlation front, and the green line shows a power-law fit \(c\cdot\tau^{p} +a\). Plotted for all measured $h_x$ values.}
    \label{fig:11}
\end{figure}

The nature of the bubble expansion is quantified by fitting the extracted boundary positions to the power-law form
\[
r(t)=a\,t^{p}+b,
\]
where \(a\), \(b\), and \(p\) are fitting parameters. The offset \(b\) accounts for the fact that the shortest experimental ramp times do not correspond exactly to \(t=0\). The uncertainty in the fitted exponent \(p\) is obtained from the covariance matrix returned by a weighted nonlinear least-squares fit. The boundary-position uncertainties are estimated from the spread of the extracted bubble front locations obtained for different initial seed spin positions and are used as weights in the fit. The reported error on \(p\) is given by the square root of the corresponding diagonal element of the covariance matrix. The complete dataset of correlation plots and fits is shown in Fig.~\ref{fig:11}.

The fitted propagation exponent stays close to \(p=1\) over the explored range of \(h_x\), indicating approximately ballistic growth of the resonant bubble within the experimentally measured timescales. At small values of \(h_x\), the bubble expands only over a small number of spins, resulting in larger uncertainties in the extracted boundary positions and consequently less stable fits. At larger values of \(h_x\), the bubble reaches the boundaries of the finite system and wraps around through the periodic boundary conditions. This leads to self-interference effects that artificially increase the apparent propagation velocity and result in fitted exponents exceeding \(p=1\).

To mitigate these effects, the fitting procedure is restricted to evolution times preceding the onset of significant self-interference. The fitting ranges were determined from the kink-density maps shown in Fig.~\ref{fig:10}, where the onset of boundary-induced effects can be identified. The corresponding truncation points are indicated by the endpoints of the bubble-boundary trajectories displayed in the correlation-function plots.

\begin{figure}
    \centering
    \includegraphics[width=\columnwidth]{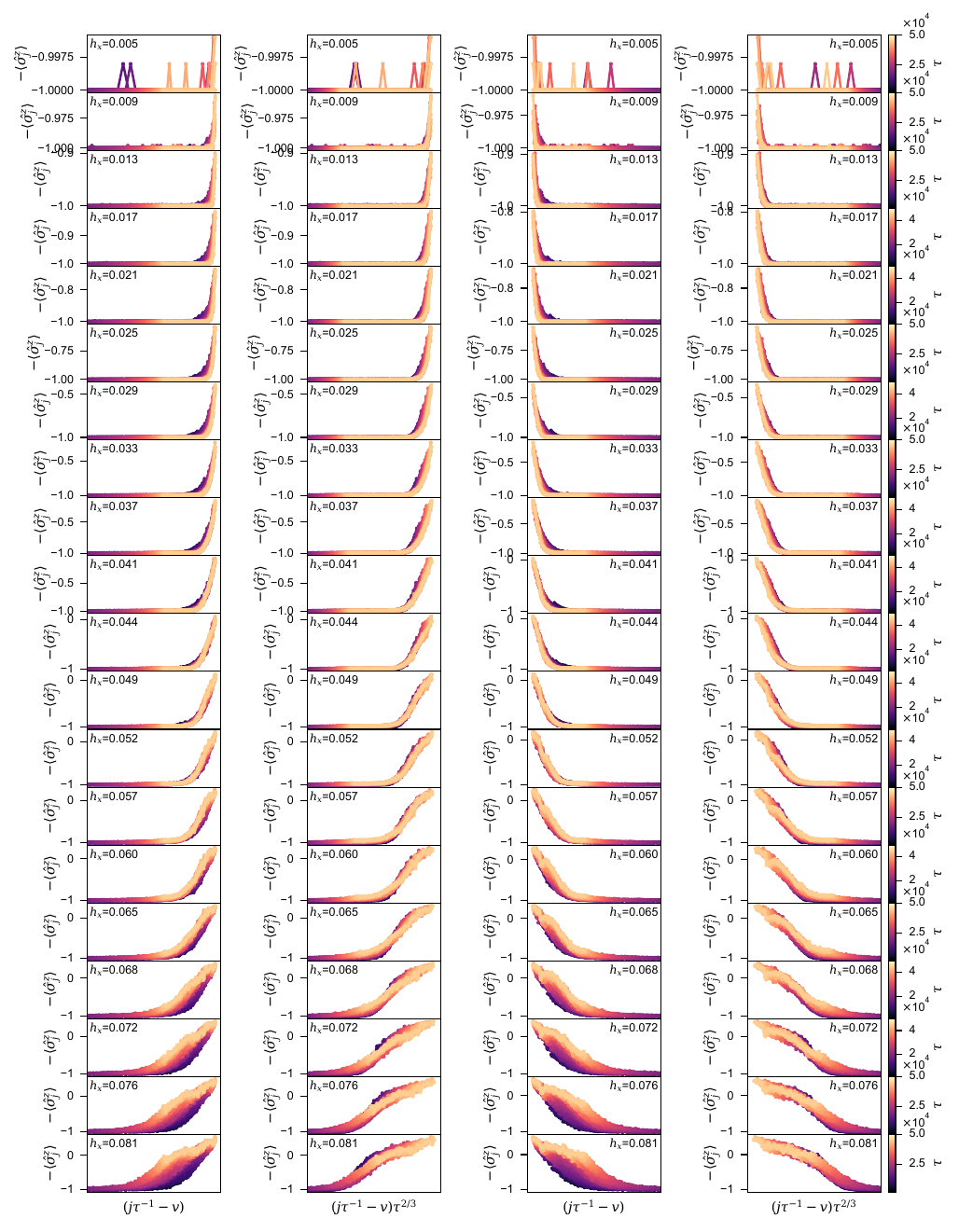}
\caption{Magnetization profiles along a one-dimensional cut for several evolution times plotted in the comoving coordinate \((j/\tau-v)\) and in the KPZ-scaled coordinate \((j/\tau-v)\tau^{2/3}\) for both the left and the right side from the initial seed location. Plotted for all measured $h_x$ values.}
    \label{fig:12}
\end{figure}

Furthermore, we provide the complete dataset of magnetization profiles obtained from the quantum simulations along a one-dimensional cut through the system for multiple evolution times and for all measured values of \(h_x\) in Fig.~\ref{fig:12}. The profiles are analyzed in both the comoving coordinate \((j/\tau - v)\) and the KPZ-rescaled coordinate \((j/\tau - v)\tau^{2/3}\), separately for the left and right sides of the initially seeded bubble.

\section{Optimization of the scaling collapse}
To further quantify the scaling of the wavefront broadening, we employed a variance-based collapse metric to determine the optimal scaling exponent numerically. Assuming a power-law scaling form, the spatial coordinate was rescaled according to \(j' = \frac{j-v\tau}{\tau^{\alpha}}\), where \(\alpha\) is the scaling exponent and \(v\) is an effective propagation velocity. For a given parameter \(\alpha\), all magnetization profiles measured at different times \(\tau\) for a fixed value of \(h_x\) were interpolated onto a common grid in the rescaled coordinate \(j'\). Denoting the interpolated profiles by \(m_i(j')\), the quality of the collapse was quantified by the mean variance across all profiles,

\begin{equation}
L(\alpha)
=
\frac{1}{N_k}
\sum_{k=1}^{N_k}
\mathrm{Var}_i
\!\left[
m_i(j'_k)
\right],
\end{equation}

where \(j'_k\) are the points of the common interpolation grid, \(N_k=100\) is the number of grid points, and \(\mathrm{Var}_i\) denotes the variance over the different times \(\tau_i\). Explicitly, 
\[
\mathrm{Var}_i
\!\left[
m_i(j'_k)
\right]
=
\frac{1}{N_\tau}
\sum_{i=1}^{N_\tau}
\left[
m_i(j'_k)-\bar{m}(j'_k)
\right]^2,
\]
with \(\bar{m}(j'_k)=\frac{1}{N_\tau}\sum_{i=1}^{N_\tau}m_i(j'_k),\), where \(N_\tau\) is the number of time points included in the analysis. The optimal scaling parameters were then obtained by numerically minimizing the cost function \(L(\alpha)\).

\begin{figure}
    \centering
    \includegraphics[width=\columnwidth]{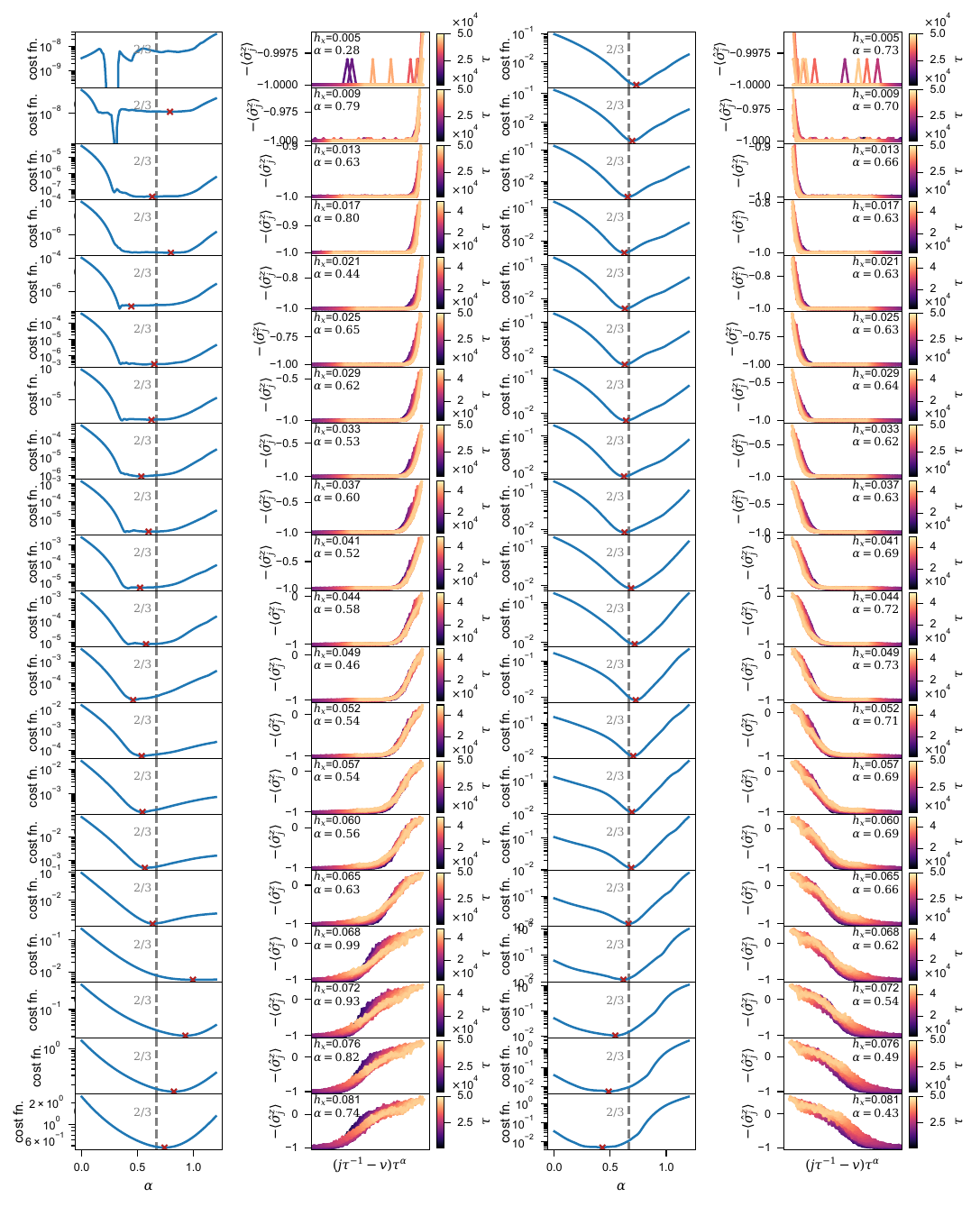}
\caption{Cost function metric values and magnetization profiles along a one-dimensional cut for several evolution times plotted in the rescaled coordinate \((j/\tau-v)\tau^{\alpha}\), where $\alpha$ represents the optimal scaling parameter determined using numerical optimization. Plotted for both the left and the right side from the initial seed location and all measured $h_x$ values.}
    \label{fig:13}
\end{figure}

The results of this analysis are shown in Fig.~\ref{fig:13}. For both sides of the seeded spin, we plot the collapse cost function as a function of the scaling exponent \(\alpha\), together with the corresponding magnetization profiles displayed in the optimally rescaled coordinate. The value \(\alpha=2/3\), corresponding to the KPZ universality class, is indicated for reference. Overall, the extracted exponents are broadly consistent with KPZ scaling. This agreement is particularly evident on the right-hand side of the seeded spin, where the cost function exhibits a well-defined minimum close to \(\alpha=2/3\) and the rescaled wavefronts collapse onto a common curve. The left-hand side shows somewhat weaker agreement, although the cost function remains relatively flat over a broad range of exponents, indicating that the data are not strongly sensitive to the precise value of \(\alpha\). Nevertheless, the optimal exponent remains compatible with the KPZ scaling within the uncertainty of the collapse procedure.

\section{Alternate stochastic circuits}

In the main text, we use a classically simulable stochastic circuit to study the long-time behavior of the system. In this Section, we explore variants of this circuit and show that it leads to the same ballistic growth in all cases. 

\begin{figure}
    \centering
    \includegraphics[width=0.5\columnwidth]{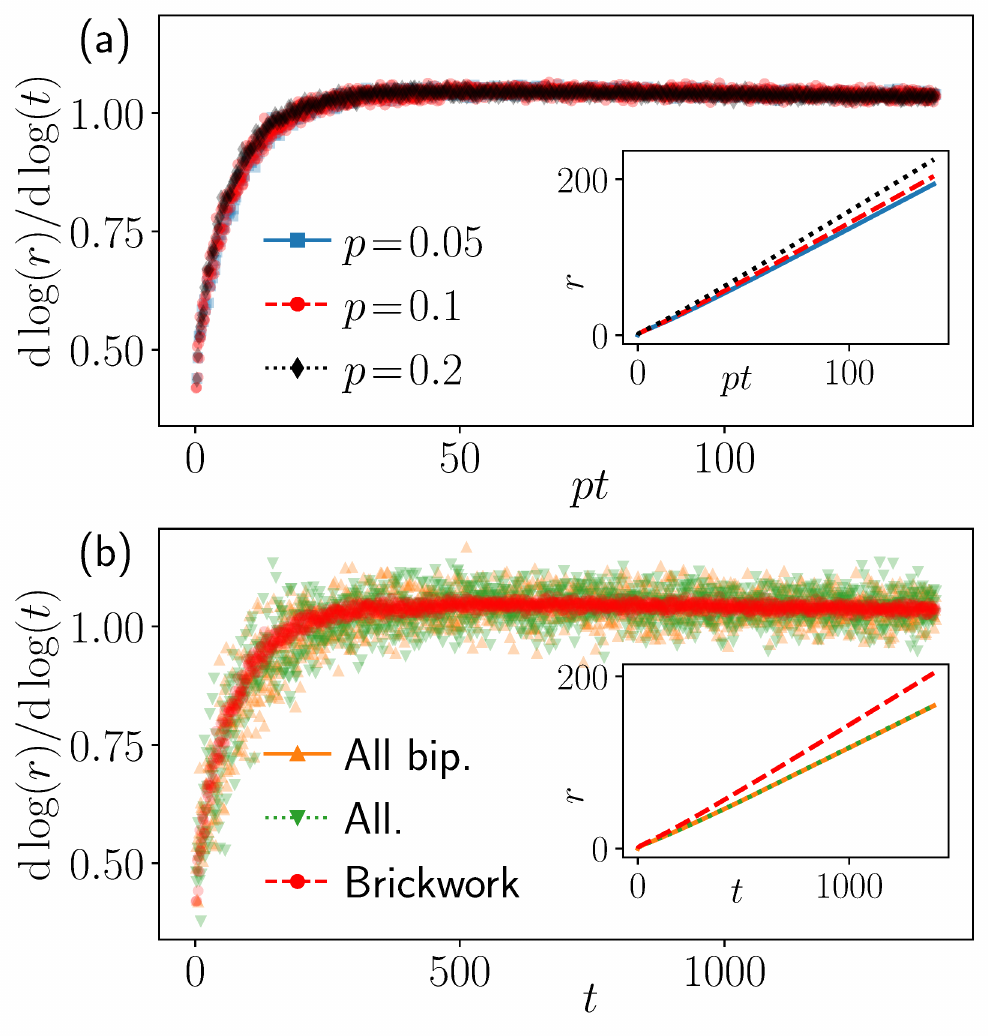}
    \caption{Comparison between the different microscopic implementations of the stochastic circuit. (a) Brickwork circuit layout alternating between the update of even and odd sites, averaged over 300 trajectories . Up to a rescaling of time, changing the flip probability $p$ has virtually no effect. (b) Comparison between the brickwork circuit (red, 3000 trajectories) and the alternative approaches where all sites can be updated together and where conflicts are resolved using either the bipartite structure of the graph (orange, 250 trajectories) or a heuristic method (green, 250 trajectories), all with $p=0.1$. In all cases a linear growth of $r$ is recovered, albeit with a slightly different slope. The all-update method leads to growth slower by a factor of approximately 1.23.}
    \label{fig:CA_all}
\end{figure}

First, Fig.~\ref{fig:CA_all}(a) shows that changing the flip probability $p$ only leads to a rescaling of time as long as $p\ll 1$. 

We also explore a non-brickwork circuit in which all sites (both even and odd) can be updated at the same step. In order to avoid violating the constraint, we need to resolve ``conflicts'' when two neighboring sites should be flipped at the same time. Finding the minimum number of flips to remove to resolve all conflicts is equivalent to finding the minimum vertex cover of a graph where each vertex is a site that should flip and edges represent conflicts between them. This is NP-hard in general. However, for bipartite graphs the minimum vertex cover
can be computed in polynomial time. Indeed, one can first find a maximum cardinality matching using Hopcroft–Karp algorithm~\cite{HopcroftKarp} and then use König's theorem to go from the matching to a vertex cover. Since the graph on which conflicts are to be resolved is a subgraph of the physical square lattice, it is bipartite and we can use this approach.

Since the bipartiteness of the graph is what motivated the brickwork approach in the first place, we also explore another method that is agnostic to it. For that, we use an heuristic algorithm to find the vertex cover that does not rely on the graph being bipartite. In practice, we turn to a local-ratio algorithm which, while usually not giving the global minimum vertex cover, is guaranteed to return a set of vertices with at most twice its cardinality~\cite{bar1985local}. All graph algorithms used relied on their implementation in the python package networkX~\cite{networkX}. The results for both of these variants are shown in Fig.~\ref{fig:CA_all}(b).

Overall, we find that all approaches lead to similar results, emphasizing that the scaling does not depend on microscopic detail but simply on the form of the constraint.

\bibliography{biblio}